\documentclass[journal,onecolumn]{IEEEtran}

\usepackage{cite}
\usepackage{amsmath,amssymb,amsfonts}
\usepackage{algorithmic}
\usepackage{graphicx}
\usepackage{textcomp}
\usepackage{xcolor}
\usepackage{boldline}
\usepackage{svg}
\usepackage{boldline}
\usepackage{multirow}
\usepackage{array}
\usepackage{hyperref}
\hypersetup{
    colorlinks=true,
    linkcolor=blue,
    citecolor=blue,
    }
\usepackage{longtable}

\usepackage[caption=false]{subfig}  
\begin{document}

\title{Investigating Brain Connectivity and Information Flow in Mental Workload Using EEG and fNIRS Integration}

\author{Mohaddese~Qaremohammadlou and Mohammad Bagher~Shamsollahi\\
\textit{Department of Electrical Engineering, Sharif University of Technology, Tehran, Iran}\\
\small Emails: 
qml.mohaddese@ee.sharif.edu, mbshams@sharif.edu}

\maketitle

\markboth{Journal}{Volume Extreme Ultraviolet Holographic Imaging}

\begin{abstract}
This study investigates brain connectivity and information flow during mental workload (MWL) by integrating electroencephalogram (EEG) and functional near-infrared spectroscopy (fNIRS) signals. Utilizing the N-back task to induce varying levels of MWL in 26 participants, we analyzed both functional and effective connectivity across 25 cortical regions derived from combined EEG and fNIRS signals. Functional connectivity was assessed using Pearson Correlation Coefficient (PCC), Phase Locking Value (PLV), and Magnitude Squared Coherence (MSC), while effective connectivity was evaluated using directed Directed Transfer Function (dDTF) and generalized Partial Directed Coherence (gPDC). Our findings reveal increased functional connectivity in frontal regions during higher MWL conditions (3-back compared to 0-back). Furthermore, effective connectivity analysis demonstrates a significant directional information flow from EEG to fNIRS, indicating a dominant influence of neural activity on hemodynamic responses. Statistical tests confirm significant differences in connectivity patterns between low and high MWL states. These results underscore the utility of EEG-fNIRS integration for characterizing brain network dynamics under varying cognitive demands and provide insights into neurovascular coupling mechanisms during mental workload.
\end{abstract}


\section{Introduction}

Mental workload (MWL) is the cognitive demand imposed on an individual during task execution and reflects the allocation of mental resources for processing, storing, and managing information. It is influenced by task complexity, cognitive capacity, and external factors such as time pressure and environmental stressors \cite{aghajani2017measuring}. MWL is a dynamic state that affects performance, attention, and decision-making. Excessive workload can lead to cognitive overload, errors, and fatigue, whereas insufficient workload may result in under-stimulation and reduced efficiency \cite{nuamah2024graph,verdiere2018detecting}.

Modern technology heightens cognitive demands, particularly in high-risk environments where operators must respond to system failures or equipment malfunctions. Quantifying the MWL is crucial for improving performance, productivity, and safety \cite{monteiro2019using}. Accurate MWL assessment has broad applications, from Brain–Computer Interfaces (BCIs) to optimizing human performance \cite{liu2017multisubject}. Elevated MWL is linked to mental fatigue and adverse health effects including immune suppression, depression, and increased risk of cardiovascular, respiratory, and musculoskeletal disorders \cite{khaksari2019effects}. Researchs have predominantly focused on prefrontal regions, given their roles in attention and working memory \cite{bourguignon2022bimodal}.

MWL detection has been extensively studied using various neurophysiological and behavioral measurements. Among these, electroencephalogram (EEG) and functional near-infrared spectroscopy (fNIRS) signals have gained prominence owing to their ability to capture neural activity changes associated with cognitive load \cite{aghajani2017measuring}. Traditional MWL detection methods include subjective assessments (e.g., NASA-TLX), behavioral metrics, and physiological measurements, with the latter being the most reliable for real-time applications \cite{chu2022optimized}. Various modalities have been explored for MWL detection including EEG, fNIRS, electrocardiogram (ECG), electromyogram (EMG), and functional Magnetic Resonance Imaging (fMRI) data. EEG is widely used because of its high temporal resolution and sensitivity to workload-related neural oscillations, whereas fNIRS provides insights into hemodynamic responses with a better spatial resolution \cite{aghajani2017measuring,verdiere2018detecting}. fMRI offers superior spatial accuracy, but is costly and lacks real-time application potential. ECG and EMG can provide indirect workload measures by assessing physiological stress responses but do not directly reflect brain activity
\cite{safari2024classification}. Regarding cost and practicality, EEG and fNIRS have emerged as optimal choices because of their portability, non-invasiveness, and ability to monitor workload in real-world settings \cite{nuamah2024graph}. EEG is preferred when precise temporal tracking of workload fluctuations is required, whereas fNIRS is advantageous for assessing cortical hemodynamic changes \cite{safari2024classification}. Combining EEG and fNIRS allows for a comprehensive assessment of MWL, leveraging the strengths of both modalities to enhance the classification accuracy \cite{aghajani2017measuring}.
Several cognitive paradigms have been utilized to investigate MWL, including the N-back task, mental arithmetic, and the Multi-Task Attribute Battery (MATB); The N-back task is a widely used working memory task that manipulates difficulty levels by adjusting the memory load, initially introduced by Kirchner in 1958 \cite{kirchner1958age}.

Brain connectivity analysis plays a crucial role in understanding information flow and connectivities between different brain regions, and is categorized into single-modality and cross-modality connectivity. The two primary approaches used for connectivity analysis are functional connectivity (FC) and effective connectivity (EC). Functional connectivity refers to the statistical dependencies between neural signals, commonly assessed using correlation-based measures, whereas effective connectivity determines the causal influence of one neural region on another one, often using Granger causality, dynamic causal modeling (DCM), or transfer entropy \cite{chan2020automated,verdiere2018detecting}.

Investigations into brain connectivity often leverage single-modalities, each providing a distinct perspective on neural activity. In the realm of EEG-based brain connectivity analysis, several studies have focused on understanding how brain networks reorganize under varying cognitive loads. These investigations predominantly use a range of functional and effective connectivity metrics to characterize neural dynamics. In the realm of functional connectivity, researchers have employed a variety of measures to explore brain network changes. For instance, a study on dynamic functional connectivity (dFC) in EEG, using Phase Locking Value (PLV) and graph theory, found that frontal-parietal connectivity was highly sensitive to changes in mental workload (MWL). Their results showed that higher MWL led to decreased characteristic path length and increased global efficiency in the Microstate D brain network, suggesting enhanced functional integration in the Theta and Alpha bands \cite{guan2022eeg}. Another study introduced a framework for task-independent MWL discrimination by fusing EEG spectral characteristics and functional connectivity, highlighting that measures like Phase Lag Index (PLI) can reveal complex, frequency- and region-dependent alterations, such as elevated frontal delta/theta power and reduced parietal alpha power, under increased cognitive demand \cite{kakkos2021eeg}.

Further investigations have delved into how specific frequency bands respond to working memory (WM) load. A study employing the weighted Phase Lag Index (wPLI) and graph theoretical metrics found that the alpha band exhibited significant reductions in connectivity and network metrics, while the high-gamma band showed a notable increase in connectivity, particularly in frontal, central, parietal, and temporal regions during difficult tasks \cite{samiei2022evaluating}. Complementing this, research using Minimum Spanning Tree (MST) analysis also quantified network reorganization, showing that increased cognitive demand shifted the theta band network towards a more decentralized, line-like topology, with significantly higher leaf fraction and lower diameter and kappa in low-load conditions \cite{nuamah2024graph}. Additionally, a study on EEG-based biometrics confirmed that functional connectivity metrics, especially PLV in the gamma band, demonstrate high discriminatory power for subject identification, with accuracy improving in higher frequency bands \cite{dutta2022functional}.

Beyond functional connectivity, some studies have leveraged effective connectivity to understand the directional flow of information. For example, a study on MWL estimation from EEG data extracted brain-effective connectivities using the directed Directed Transfer Function (dDTF). The results demonstrated that these effective connectivity patterns could effectively distinguish between high and low MWL levels, highlighting specific directed neural activity flows that differentiate cognitive states \cite{safari2024classification}. This concept of cognitive reorganization has been a central theme, as another study found statistically significant reductions in clustering coefficient and small-worldness metrics with higher workload, suggesting that the brain network rewired for more efficient long-range communication to meet increased cognitive demands \cite{dimitrakopoulos2017task}. These findings are further supported by a study on simulated flight experiments, which showed distinct alterations in functional brain networks and differing topological patterns in beta band global efficiency between 2D and 3D interfaces under varying MWL levels \cite{kakkos2019mental}.

In the analysis of brain connectivity within fNIRS studies, various investigations have been conducted. For instance, a study notably demonstrated that increased mental workload, induced by a pattern recognition task, leads to a robust enhancement in prefrontal cortex (PFC) functional connectivity, as quantified by metrics such as network density, clustering coefficient, and global efficiency. These findings underscore the dynamic nature of brain networks in response to cognitive demands and emphasize fNIRS's potential as a tool for investigating real-time changes in brain function \cite{racz2017increased}. another study focusing on mental fatigue compared brain connectivity across five frequency bands, mental fatigue induces distinct changes in brain functional connectivity across three levels of fatigue (non-fatigue, moderate, and severe). As fatigue progressed to a severe state, in higher frequency bands (0.6-2.0 Hz), overall connectivity decreased, especially between the PFC and other regions. The PFC, Frontal Eye Field (FEF), Premotor Cortex (PMC), and Supplementary Motor Area (SMA) consistently showed the most pronounced changes in connectivity and network characteristics, reflecting their involvement in cognitive processes affected by fatigue \cite{peng2022functional}. Complementing these, a new methodological approach was introduced for the analysis of fNIRS connectivity by describing a differential symmetry index (DSI) to quantify the incommon physiological information between functional connectivity networks of oxy-haemoglobin (HbO) and deoxy-haemoglobin (HbR). Their work highlighted that for sparse FC networks, it is strongly recommended to analyze both hemoglobin species to avoid potentially misleading inferences from incomplete brain network information, providing a crucial guideline for fNIRS connectivity analysis \cite{montero2018estimating}.

The relationships between physiological signals have been studied extensively using various methods. However, specific cross-modality connectivity studies for mental workload remain scarce, leading researchers to draw insights from related fields. For instance, the coupling between EEG and fNIRS has been evaluated by analyzing the simultaneous variations in HbO and HbR concentrations, along with changes in the alpha and beta frequency bands of EEG signals \cite{lachert2017coupling}. Brain–heart interactions during injection procedures have been investigated by comparing spectral power changes in EEG within the 33–36 Hz frequency band and mean heart rate (HR) values at different injection depths, where Granger causality analysis indicated a stronger directional connection from the brain to the heart, which increased with injection depth \cite{won2019alteration}. Another study examined the relationship between EEG and ECG at varying heart rates and respiration rates using the Magnitude Squared Coherence (MSC) metric, demonstrating that EEG signal information could be extracted from ECG signals by analyzing power spectral density \cite{singh2012estimation}.
In Parkinson’s disease patients, the EEG-ECG connection was found to be weaker than in healthy individuals, particularly when analyzing the relationship between alpha and gamma oscillations and heart rate fluctuations \cite{candia2023coupling}. Another study explored EEG-EMG interactions in stroke patients compared to healthy individuals using transfer entropy analysis during motor tasks (e.g., grasping with the left and right hand and elbow flexion). The results showed higher bidirectional transfer entropy (EMG-to-EEG and EEG-to-EMG) in patients with stroke, with the strongest connectivity observed in the beta frequency band (15–35 Hz) during upper limb movement. Furthermore, EMG-to-EEG coupling was found to be stronger in the affected limb, whereas EEG-to-EMG connectivity was lower \cite{gao2018electroencephalogram}. In Alzheimer’s disease (AD), EEG-fNIRS connectivity reveals weakened alpha- and beta-band connectivity in the orbitofrontal and parietal regions, highlighting the potential of multimodal fusion for early AD detection \cite{li2019dynamic}.
Another study applied nonlinear Granger causality with NARX modeling to analyze cardiovascular signal interactions, and their results showed that respiration strongly influences the heart rate and blood oxygen concentration, demonstrating the effectiveness of nonlinear methods in physiological connectivity analysis \cite{bahrami2023investigating}. In EEG-PPG Fusion, explores the interaction between EEG and photoplethysmogram (PPG) to assess deception detection through effective connectivity analysis. The researchers introduced a wavelet-based fusion approach for EEG and PPG signals, which was validated on 41 subjects who participated in an interview-based deception scenario. Using generalized Partial Directed Coherence (gPDC) and dDTF, they constructed effective connectivity networks to analyze the information flow between the brain and autonomic nervous system responses. The findings revealed distinct connectivity patterns between guilty and innocent individuals, with increased connectivity in specific brain regions in those engaging in deception. This study highlights the role of brain-autonomic coupling in cognitive and emotional processes, emphasizing the importance of multimodal physiological connectivity in forensic and cognitive research \cite{daneshi2020eeg}.

This study builds on previous EEG-PPG effective connectivity research \cite{daneshi2020eeg}, which highlighted the advantages of multimodal integration for cognitive state monitoring. Expanding on these findings, we investigated EEG-fNIRS coupling and information flow under mental workload conditions. Our approach involved EEG-fNIRS integration across 25 cortical regions (R1–R25), constructing functional connectivity matrices (PCC, PLV, MSC), and effective connectivity metrics (dDTF, gPDC) to assess network interactions and directional information flow. We compared connectivity patterns between low- and high-workload states and performed statistical analysis across 26 subjects to identify significant MWL-related connectivity changes. This study aimed to enhance EEG-fNIRS connectivity analysis and contribute to real-time MWL monitoring applications in cognitive neuroscience and human performance research.

\section{MATERIALS AND METHODES}

\subsection{Experiment}
In this study, an open-source N-back dataset collected at the Technical University of Berlin was utilized \cite{shin2018simultaneous}. This dataset comprises three sessions, each including three series corresponding to the 0-back, 2-back, and 3-back tasks. Within each series, a 2-second instructional phase introduced the specific task type ($n=0,2,3$), followed by a 40-second active task period and a subsequent 20-second rest period. A short, 250 ms sound marks the beginning and end of each active period. Upon completion of the active period, the monitor displays the word "STOP" for a duration of 1 s, subsequently transitioning into the rest period where a fixation cross is shown on the screen. Each active session included 20 trials, each of which was considered a discrete stimulus. During each trial, a randomly generated single-digit number appears on the monitor every 2 seconds, remaining visible for 0.5 seconds before a fixation cross occupies the screen for the remaining 1.5 seconds. Targets appear with a 30$\%$ probability, with the remaining 70$\%$ representing nontargets. In the 0-back task, participants were instructed to press the designated "target" button (number 7) or the "non-target" button (number 8) in response to a specific predetermined character, serving as a preliminary measure to confirm their familiarity with the task protocol. In the 2-back and 3-back conditions, participants press the "target" button if the presented digit matches the one shown 2 or 3 trials prior, respectively.

\begin{figure*}
    \centering
 \hspace{-0.1cm}
  \includegraphics[scale=0.45]{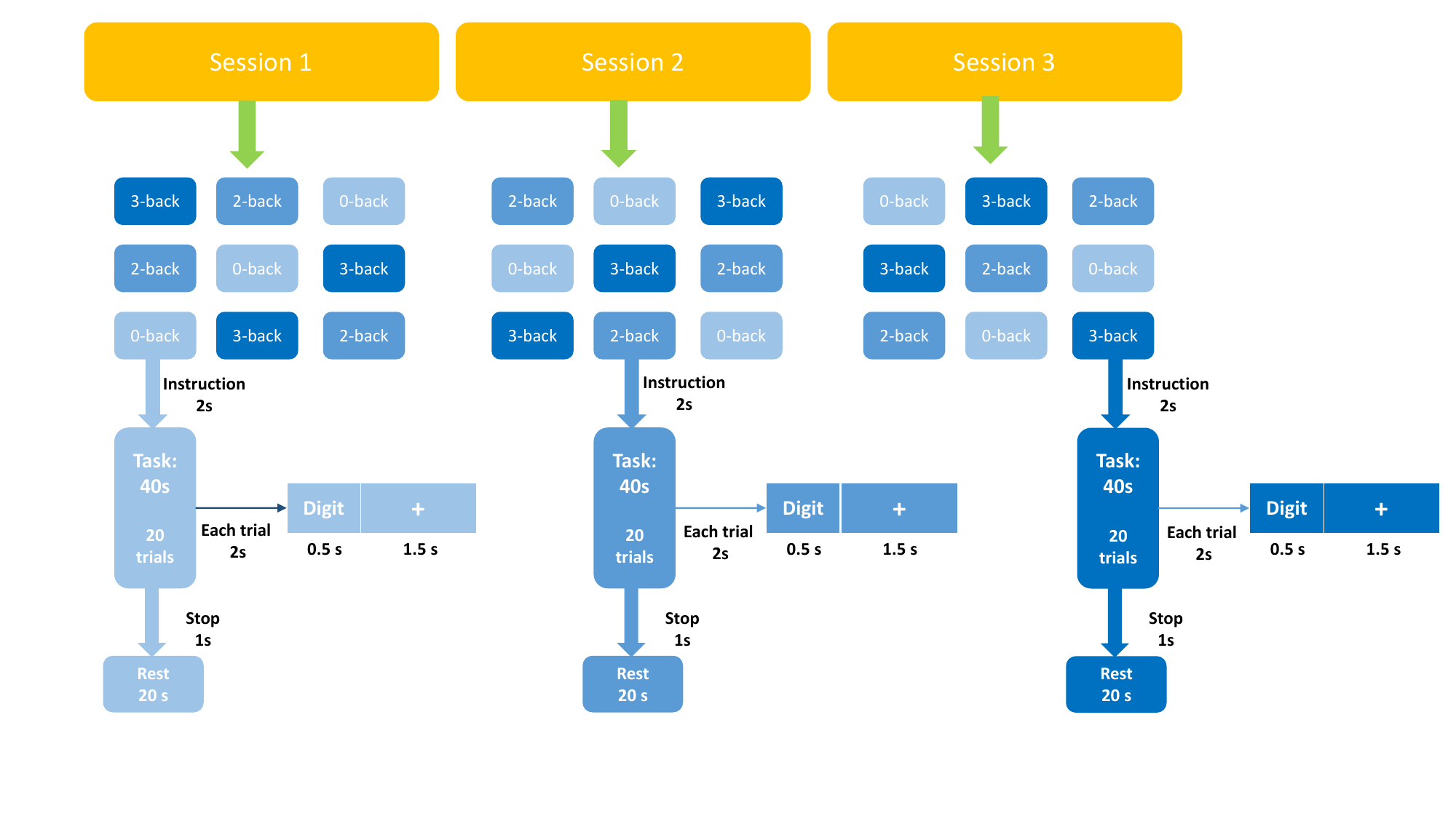}
    \vspace{-0.5cm}
    \caption{Experiment paradigm of the N-back task. Each session consists of three series for 0-back, 2-back, and 3-back tasks, presented in a randomized order. Each series includes a 2-second instruction phase indicating the task type, a 40-second active task period with 20 trials, and a 20-second rest period. Each trial within the active task period is 2 seconds long, comprising a 0.5-second digit presentation and a 1.5-second fixation cross. Auditory cues signal the start and end of the active task period, and the word "STOP" is displayed for 1 second after each active period, followed by the rest period with a fixation cross \cite{qaremohammadlou2024mental}.}
    \label{fig1}
\end{figure*}

Throughout the rest period, a fixation cross remained on the screen for 20 s, allowing participants to relax and minimize eye movement, thereby supporting the brain's return to baseline activity. For each n-back task condition, 180 trials were conducted, calculated as 20 trials per series across three series and three sessions  (Fig.~\ref{fig1}).

\subsection{Data Acquisition and Participants}
The study included twenty-six healthy participants (9 male and 17 female; mean age = 26 $\pm$ 3 years). Data were acquired concurrently using EEG and  fNIRS. EEG data were sampled at a frequency of 1000 Hz across 30 channels arranged according to the international 10-5 electrode placement system, with TP9 and TP10 electrodes serving as reference and ground electrodes, respectively. The fNIRS data were recorded at a sampling rate of 10.4 Hz across 36 channels \cite{shin2018simultaneous}. 

\subsection{Preprocessing} EEG and fNIRS data underwent separate preprocessing, utilizing filtering techniques and methods specific to each modality:

\subsubsection{EEG Preprocessing:}The EEG data sampling rate was initially reduced from 1000 Hz to 200 Hz. Subsequently, a sixth-order Butterworth bandpass filter with a frequency range of 1–40 Hz was applied. Eye movement artifacts were removed using the Adaptive Artifact Removal (AAR) toolbox in conjunction with the iWASOBI method \cite{gomez2007automatic}.

\subsubsection{fNIRS Preprocessing:} Raw fNIRS data were converted to reflect changes in deoxygenated hemoglobin concentration ($\Delta$HbR or DEOXY) and oxygenated hemoglobin concentration ($\Delta$HbO or OXY) using the modified Beer-Lambert law (MBLL) \cite{hwang2014evaluation,blankertz2010berlin}. The sampling rate was adjusted from 10.4 Hz to 10 Hz. Owing to the low intrinsic frequency of this signal (0.017 Hz) and the mitigation of high-frequency components and systemic physiological noise (e.g., cardiac and respiratory), a sixth-order Butterworth low-pass filter (cutoff frequency: 0.2 Hz) and a third-order Savitzky-Golay (SG) filter were applied \cite{khanam2022statistical}. 

Following preprocessing, the EEG and fNIRS signals were segmented based on dataset labels to create target and non-target samples for each task type (0-back, 2-back, and 3-back). 
Smaller segments of EEG and fNIRS signals have been extracted in various ways in the literature: a time window of 0.1 s before stimulation and 1 s after stimulation \cite{salimi2022predicting}, 500 ms before stimulation and 6 s after stimulation \cite{cao2022eeg}, 1 s before stimulation and 3 s after stimulation \cite{samiei2022evaluating}, and time windows of 2 to 5 s and 1 s \cite{saadati2020convolutional,khan2020hybrid, saadati2019mental}.

Based on  outcomes from various segment lengths, the selected segment for this study was 0.5 s before and 10 s after stimulation for EEG data and 5 s before and 30 s after stimulation for fNIRS data. The extended segment length for the fNIRS data accommodates the delayed hemodynamic response following stimulation.
A total of 180 samples per class were used across the 0-back, 2-back, and 3-back conditions for each data type (EEG, OXY, and DEOXY) \cite{qaremohammadlou2024mental}. The relationship between EEG and fNIRS signals was analyzed by extracting the upper envelope of the 10 Hz wavelet component for EEG data and the 1 Hz component for fNIRS data \cite{daneshi2020eeg}. 

\begin{figure*}

\centering
\subfloat[EEG-regions.]{
    \label{figure2a}
    \includegraphics[width=0.3\textwidth]{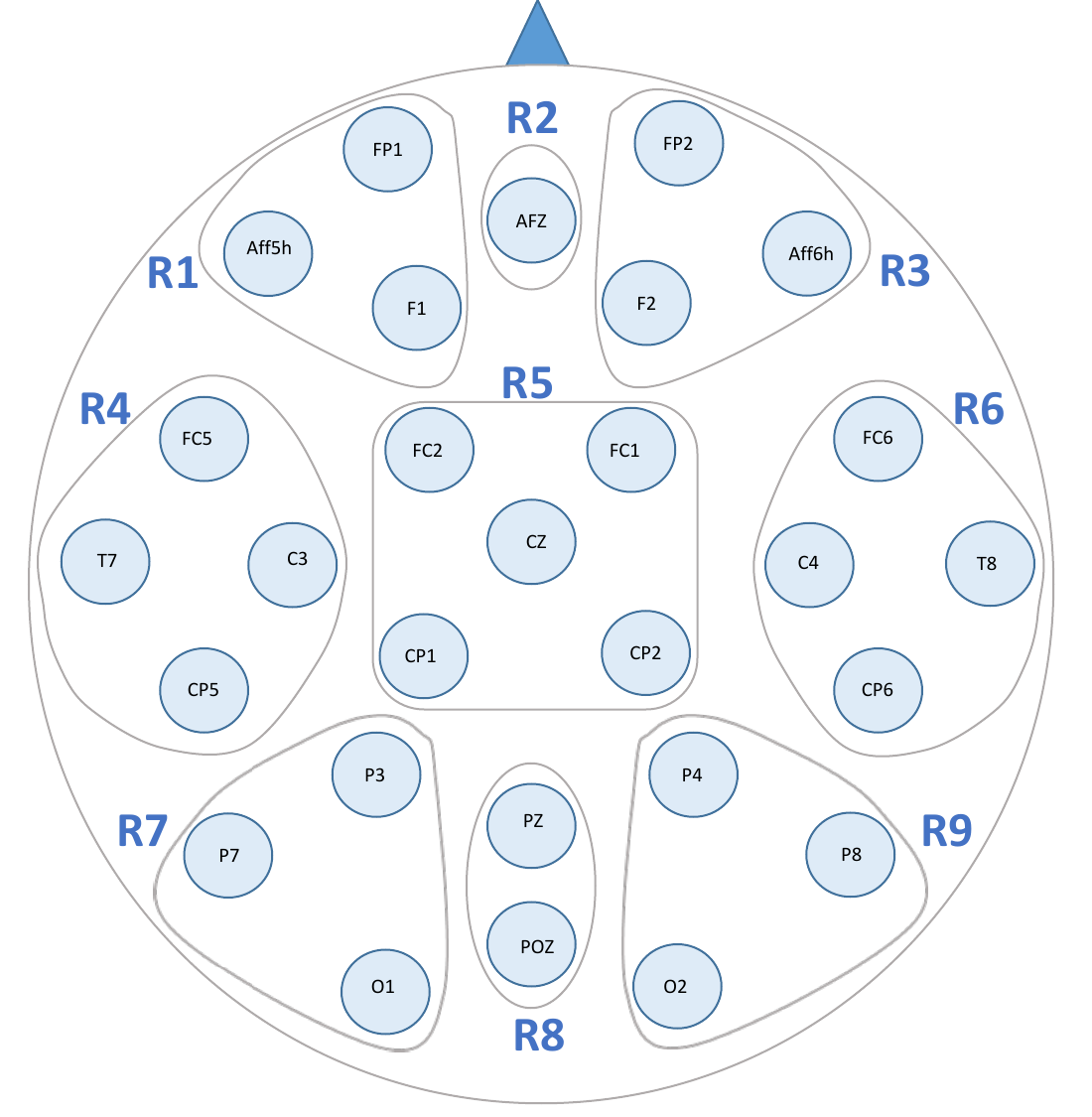}
}
\hfill
\subfloat[OXY-regions.]{
    \label{figure2b}
    \includegraphics[width=0.3\textwidth]{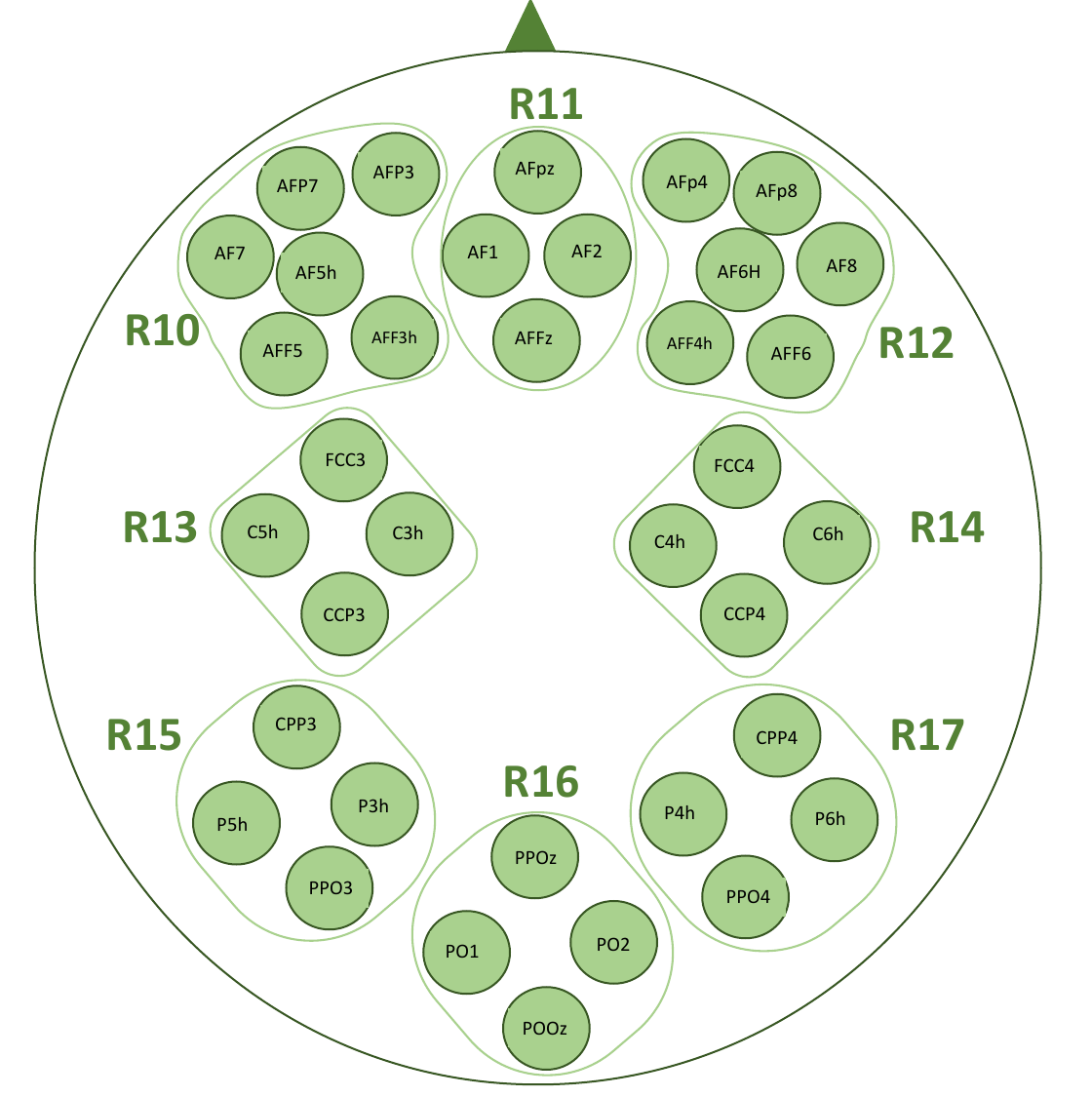}
}
\hfill
\subfloat[DEOXY-regions.]{
    \label{figure2c}
    \includegraphics[width=0.3\textwidth]{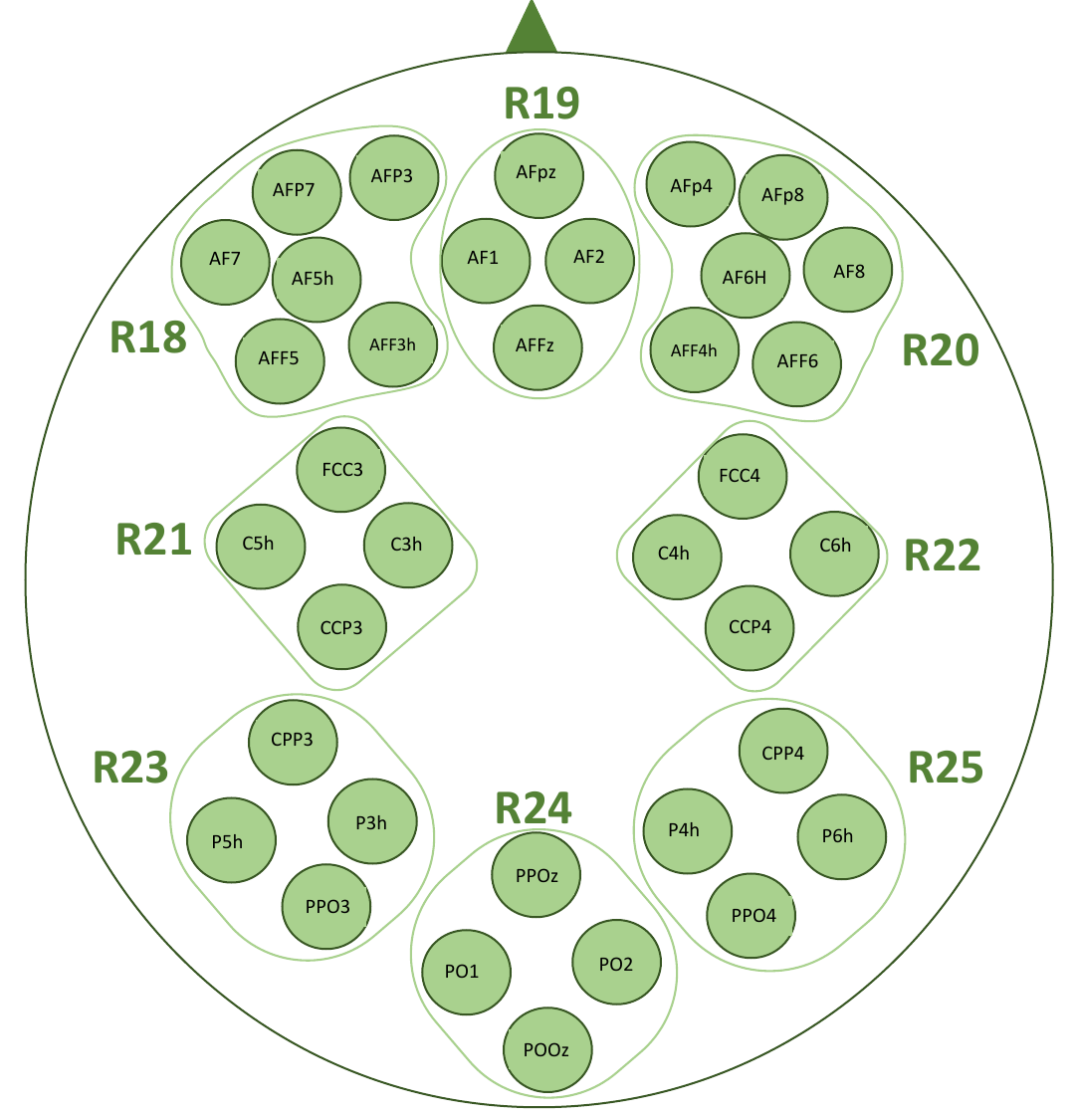}
}

\caption{Region segmentation method for EEG and fNIRS datasets. (a) EEG-regions: Depicts the nine regions (R1-R9) defined for EEG data, based on the international 10-5 electrode placement system. (b) OXY-regions: Shows the eight regions (R10-R17) defined for oxygenated hemoglobin (OXY) fNIRS data. (c) DEOXY-regions: Illustrates the eight regions (R18-R25) defined for deoxygenated hemoglobin (DEOXY) fNIRS data. Each region represents an average signal from the channels within that area for connectivity analysis.}

\label{fig2}
\end{figure*}

\begin{figure*}[h!]
    \centering
  \includegraphics[scale=0.55]{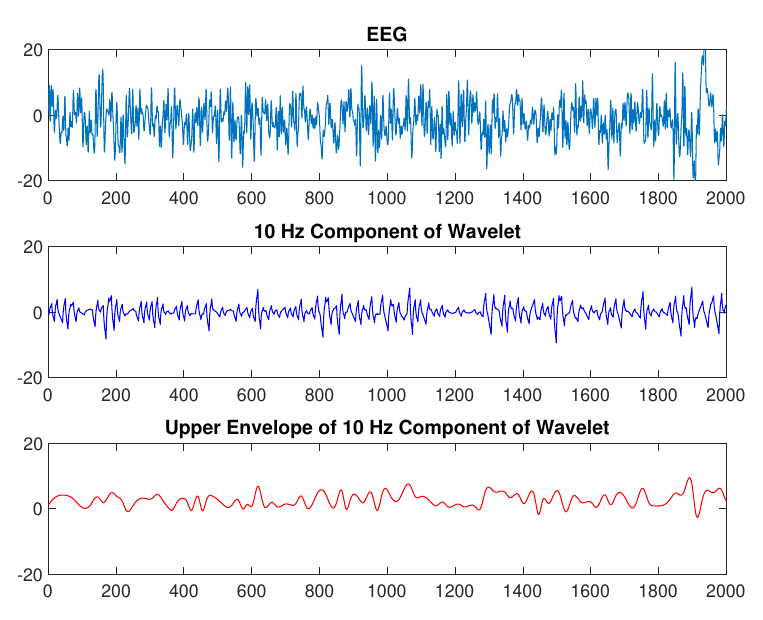}
    \vspace{0.05cm}
    \caption{Example of signal preprocessing for EEG data. Top panel: The preprocessed EEG signal from a single channel. Middle panel: The 10 Hz wavelet component extracted from the EEG signal using wavelet transform with a db2 mother wavelet. Bottom panel: The upper envelope of the 10 Hz wavelet component, which is used for further functional and effective connectivity analysis.}
    \label{fig3}
\end{figure*}

 As shown in Fig.~\ref{fig2}, the average of the channels within each region was calculated for the EEG, OXY, and DEOXY data, with nine regions defined for the EEG data and eight regions specified for each fNIRS data type. Owing to the differing dynamics of the EEG and fNIRS signals, a wavelet transform with a db2 mother wavelet was applied to both the signals. Ultimately, the 1 Hz wavelet component of the fNIRS signal and the upper envelope of the 10 Hz wavelet component of the EEG signal were extracted. Fig.~\ref{fig3} illustrates an example of the preprocessed EEG signal, 10 Hz wavelet component, and upper envelope of the 10 Hz wavelet component. After normalizing the data for each channel, low-frequency data comprising 25 channels were available and used to calculate effective and functional brain connectivities.

 Functional connectivity metrics were compared across two workload levels, 0-back (low level) and 3-back (high level), to examine changes in functional connectivity between EEG and fNIRS signals. The metrics used included PCC, PLV, and MSC, each calculated for the 0-back and 3-back levels. To investigate the effective connectivity between the EEG and fNIRS signals, effective connectivity metrics were calculated for two workload levels: 0-back and 3-back. The metrics used included dDTF and gPDC, and calculations were performed using the \href{https://github.com/sccn/SIFT }{SIFT toolbox} \cite{sift2011,Sift2014}. For each subject, an MVAR model was first fitted to the data, and the effective connectivity metrics were then computed.
In the SIFT toolbox, as illustrated in Table~\ref{tab1}, the vertical axes of the connectivity matrices represent the flow of information "from" a channel, while the horizontal axes indicate the flow of information "to" a channel.

\begin{table}[h!]
\renewcommand{\arraystretch}{1.5}
    \centering
    \caption{Connectivity Matrix} 
    \label{tab1}
    \begin{tabular}{|c|c|c|c|}
        \cline{2-4} 
        \multicolumn{1}{c|}{} & \multicolumn{3}{c|}{From (output)} \\
        \cline{1-4} 
        \multirow{3}{*}{\rotatebox{90}{To (input)}} & EEG & OXY & DEOXY \\
        \cline{2-4} 
        & OXY & & \\
        \cline{2-4} 
        & DEOXY & & \\
        \cline{1-4} 
    \end{tabular}
\end{table}

\section{Results}
The results of brain connectivities are detailed below and include both functional and effective brain connectivity.
\subsection{Results of Functional Connectivity from EEG and fNIRS Signals Coupling}
The matrices presented for each functional connectivity metric (PCC, PLV, and MSC) at the 0-back and 3-back workload levels corresponded to all the participants in the study.
\subsubsection{PCC:}
\begin{figure*}[h!]
    \centering
    \subfloat[0-back.]{\includegraphics[width=0.45\textwidth]{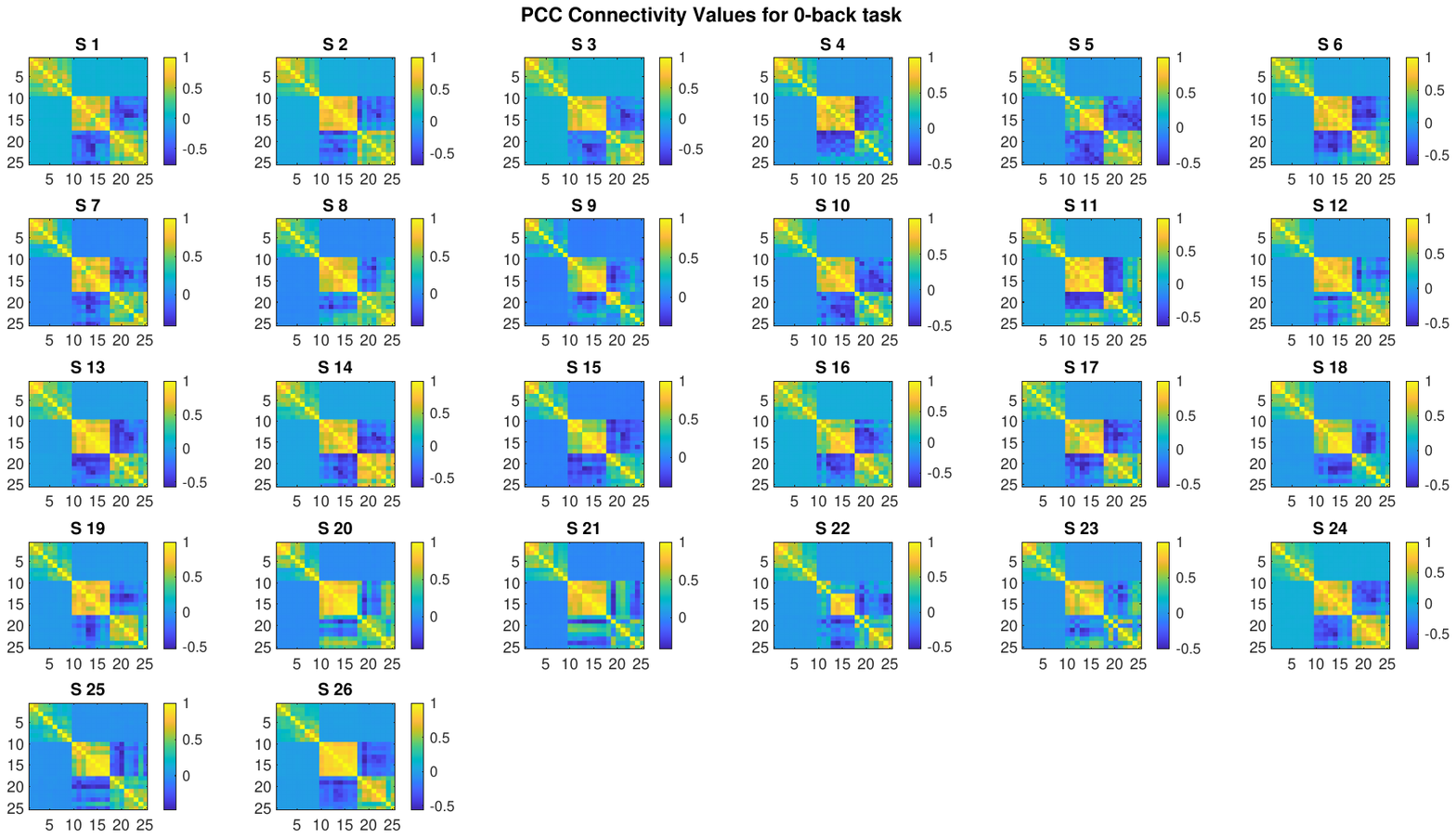}}
    \hfill
    \subfloat[3-back.]{\includegraphics[width=0.45\textwidth]{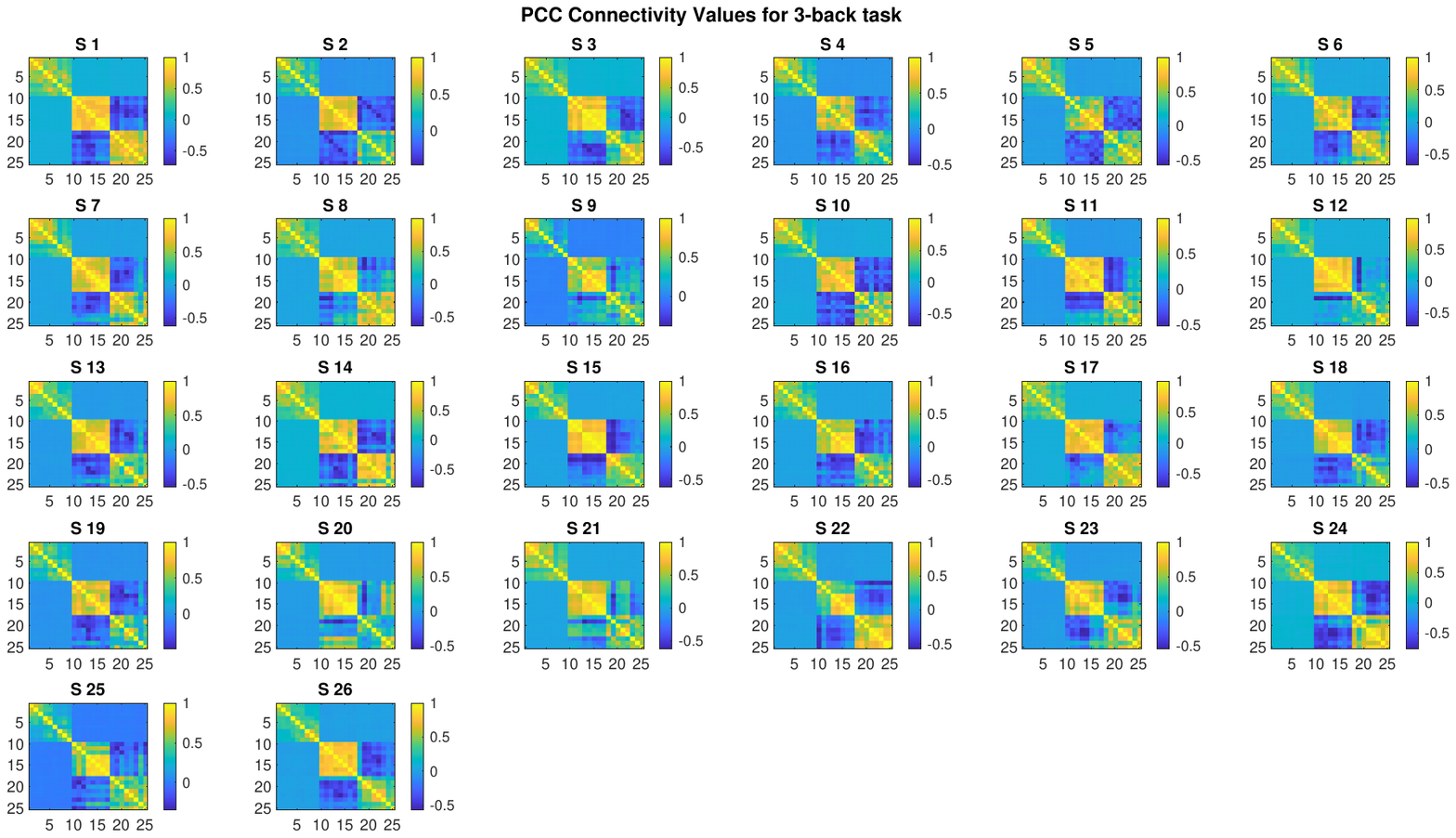}}
    \caption{Pearson Correlation Coefficient (PCC) connectivity matrices for the 0-back (a) and 3-back (b) tasks. Each matrix represents the functional connectivity between 25 brain regions (R1-R25) for 26 individual subjects (S1-S26), with data derived from combined EEG and fNIRS signals. The color scale, ranging from -0.5 to 1, indicates the strength of the correlation. Higher values (yellow/red) signify stronger positive correlations, while lower values (blue) denote weaker or negative correlations. Comparing the two sub-figures, notice the distinct changes in connectivity patterns under the higher cognitive workload of the 3-back task.}
    \label{fig4_5}
\end{figure*}

Based on Figure ~\ref{fig4_5}, which displays the PCC values for the 0-back and 3-back tasks, an overall increase in brain connectivity was observed in the frontal regions of the head (R1, R2, R3). A comparison of the connectivity values between the 0-back and 3-back tasks revealed that connectivity was higher in the 3-back task. Furthermore, in regions R10, R11, R12, ... R17, corresponding to OXY data, increased connectivity was observed.

\subsubsection{PLV:}
\begin{figure*}[h!]
    \centering
    \subfloat[0-back.]{\includegraphics[width=0.45\textwidth]{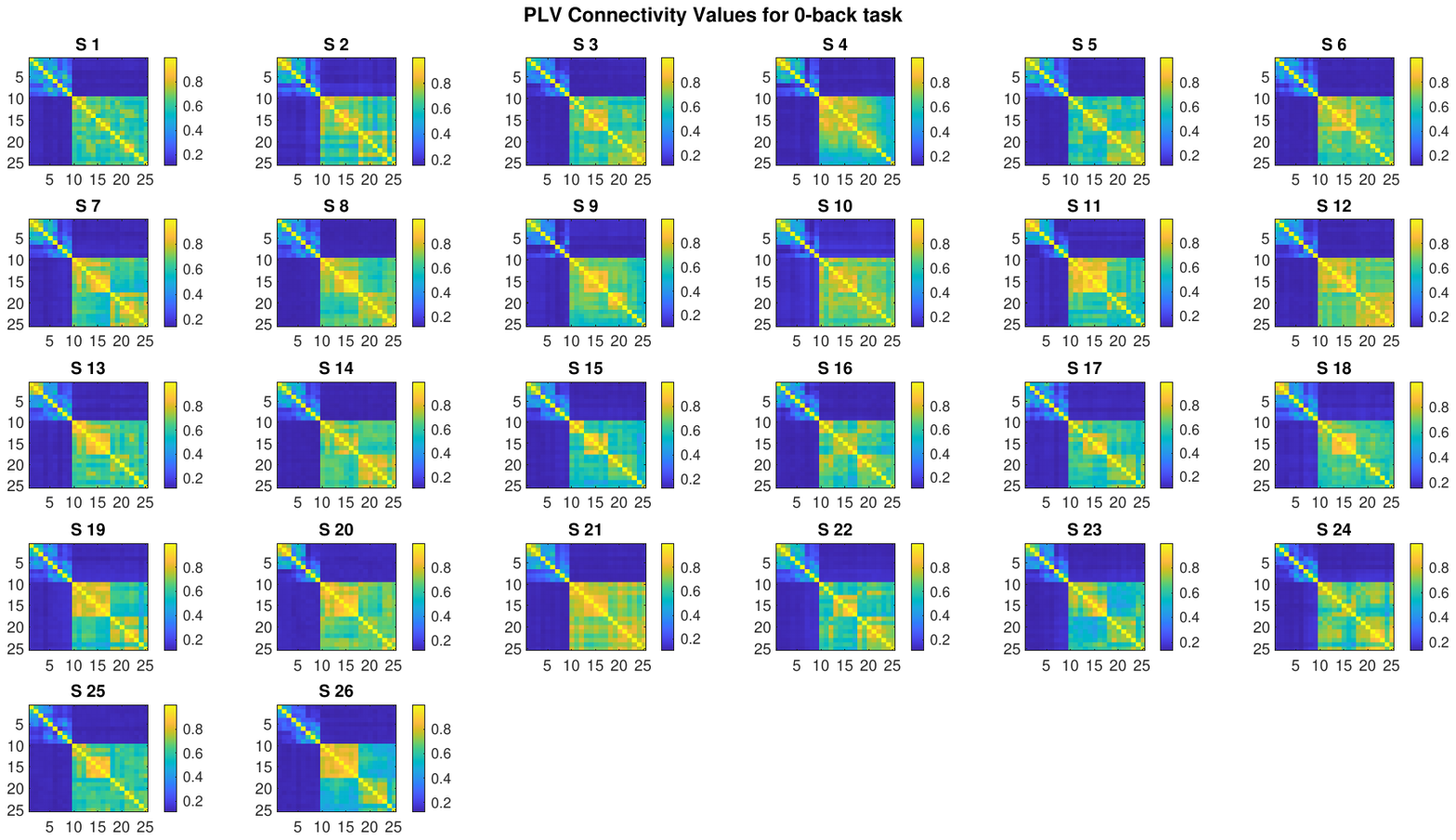}}
    \hfill
    \subfloat[3-back.]{\includegraphics[width=0.45\textwidth]{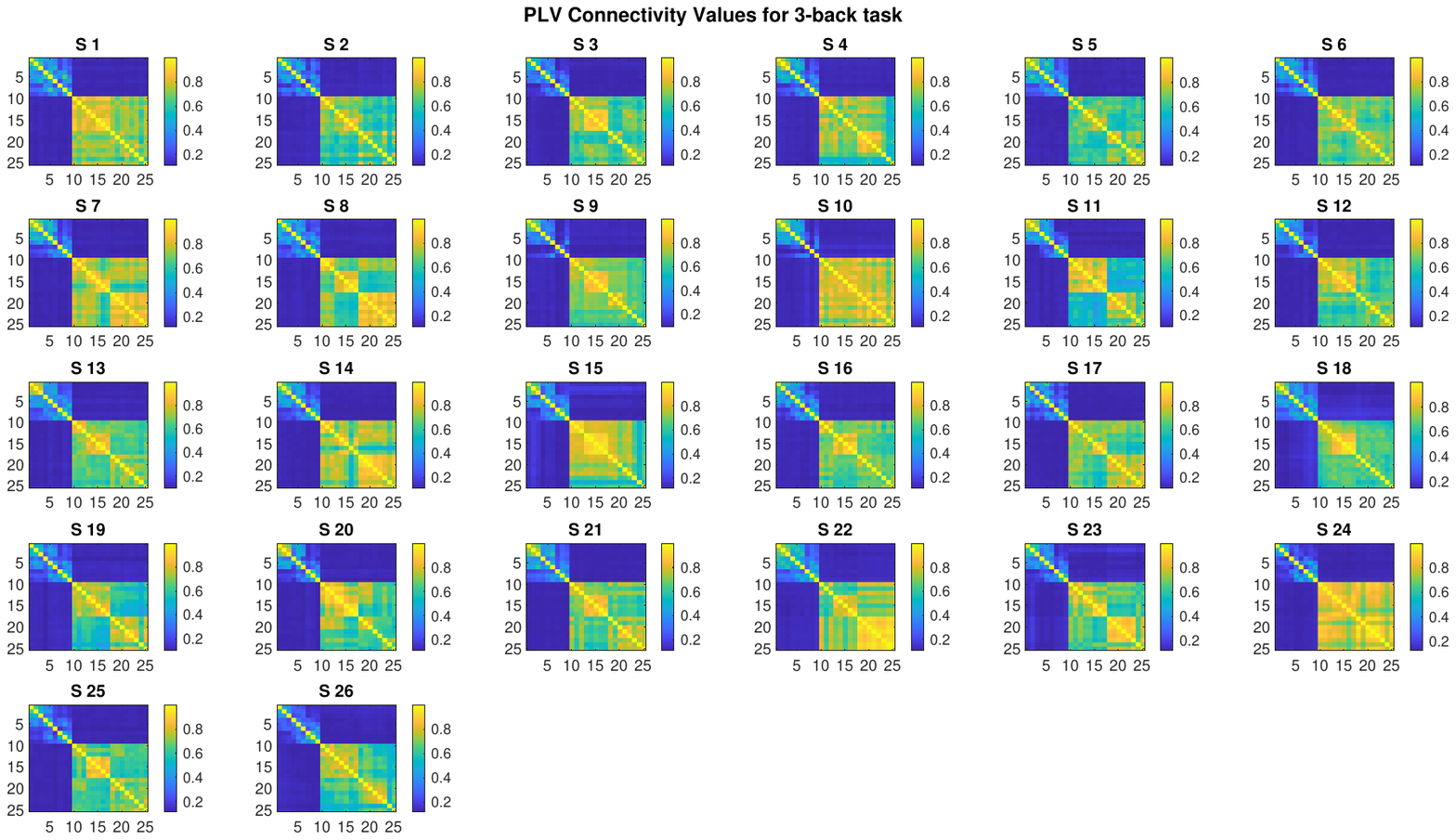}}
    \caption{Phase Locking Value (PLV) connectivity matrices for the 0-back (a) and 3-back (b) tasks. Each matrix represents the functional connectivity between 25 brain regions (R1-R25) for 26 individual subjects (S1-S26), with data derived from combined EEG and fNIRS signals. The color scale, ranging from 0.2 to 0.8, indicates the strength of phase synchronization. Higher values (yellow/red) signify stronger phase locking between regions. Comparing the two sub-figures, notice the distinct changes in phase synchronization patterns under the higher mental workload of the 3-back task.}
    \label{fig6_7}
\end{figure*}

Based on Figure ~\ref{fig6_7}, which displays the PLV values for the 0-back and 3-back tasks, respectively, an overall increase in brain connectivity was observed in the frontal regions of the head (R1, R2, R3). A comparison of the connectivity values between the 0-back and 3-back tasks indicated that connectivity was higher in the 3-back task. Additionally, greater connectivity was observed in regions associated with fNIRS data than in those related to EEG data.

\subsubsection{MSC:}
\begin{figure*}[h!]
    \centering
    \subfloat[0-back.]{\includegraphics[width=0.45\textwidth]{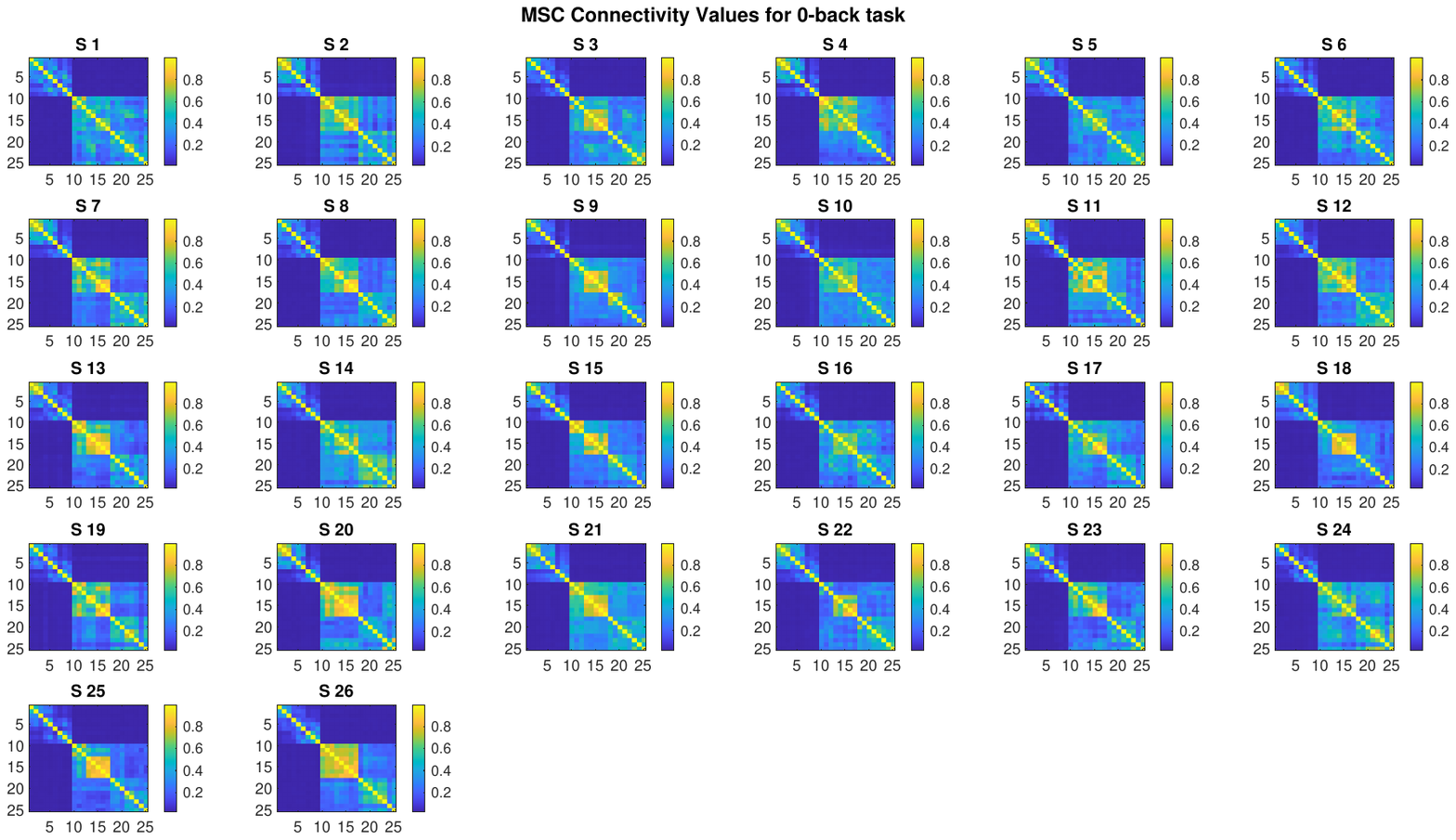}}
    \hfill
    \subfloat[3-back.]{\includegraphics[width=0.45\textwidth]{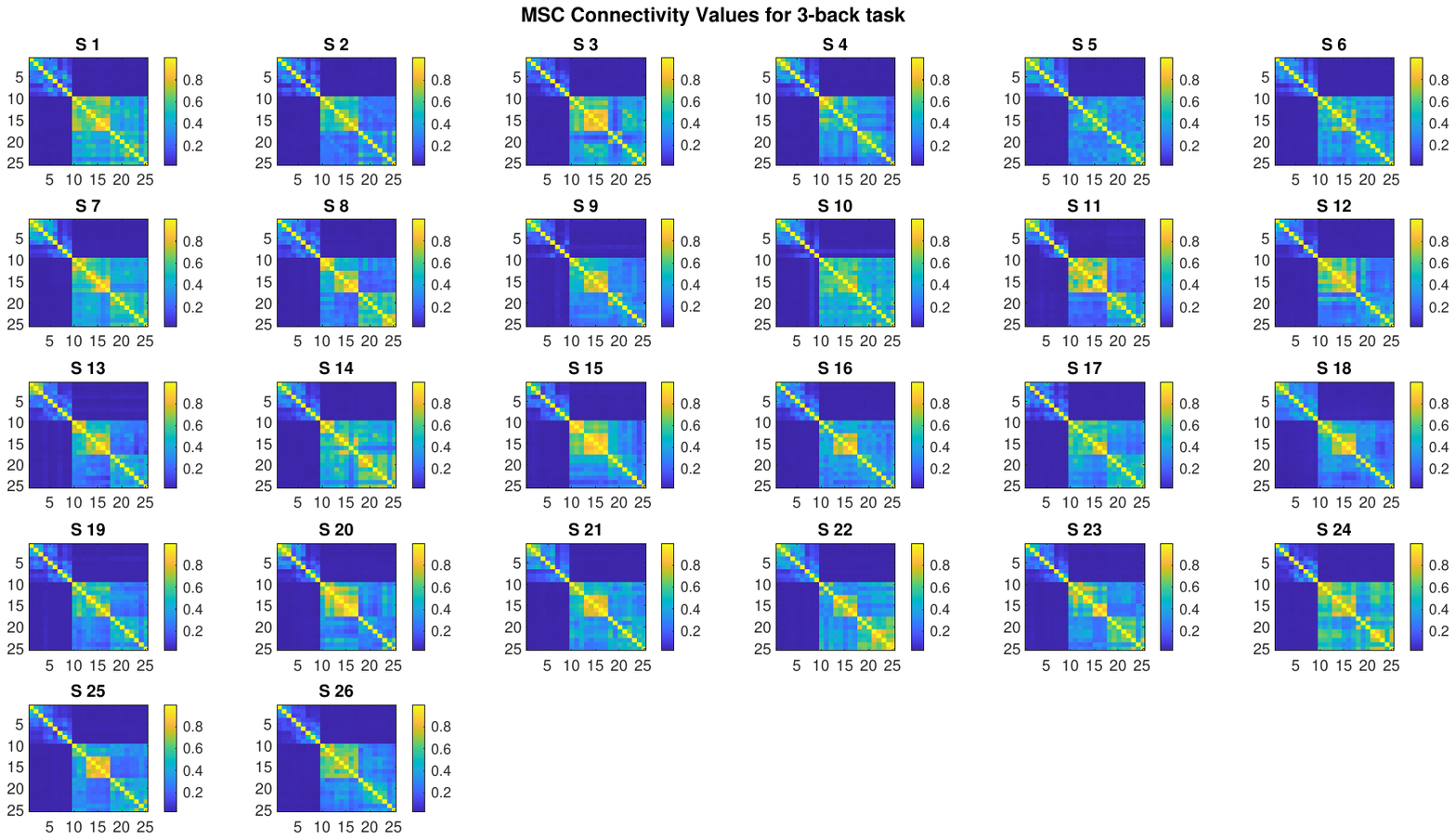}}
    \caption{Magnitude Squared Coherence (MSC) connectivity matrices for the 0-back (a) and 3-back (b) tasks. Each matrix represents the functional connectivity between 25 brain regions (R1-R25) for 26 individual subjects (S1-S26), with data derived from combined EEG and fNIRS signals. The color scale, ranging from 0.2 to 0.8, indicates the strength of coherence. Higher values (yellow/red) signify a stronger linear dependency in the frequency domain between regions. Comparing the two sub-figures, notice the distinct alterations in frequency domain coherence due to the increased mental workload of the 3-back task.}
    \label{fig8_9}
\end{figure*}

Based on Figure ~\ref{fig8_9}, which displays the MSC values for the 0-back and 3-back tasks, respectively, an overall increase in brain connectivity was observed in the frontal regions of the head (R1, R2, R3). A comparison of the connectivity values between the 0-back and 3-back tasks indicated that connectivity was higher in the 3-back task. Additionally, greater connectivity was observed in regions associated with the OXY data.

\subsection{Results of Effective Connectivity from EEG and fNIRS Signals Coupling}
\begin{figure*}
    \centering
    \includegraphics[scale=0.4]{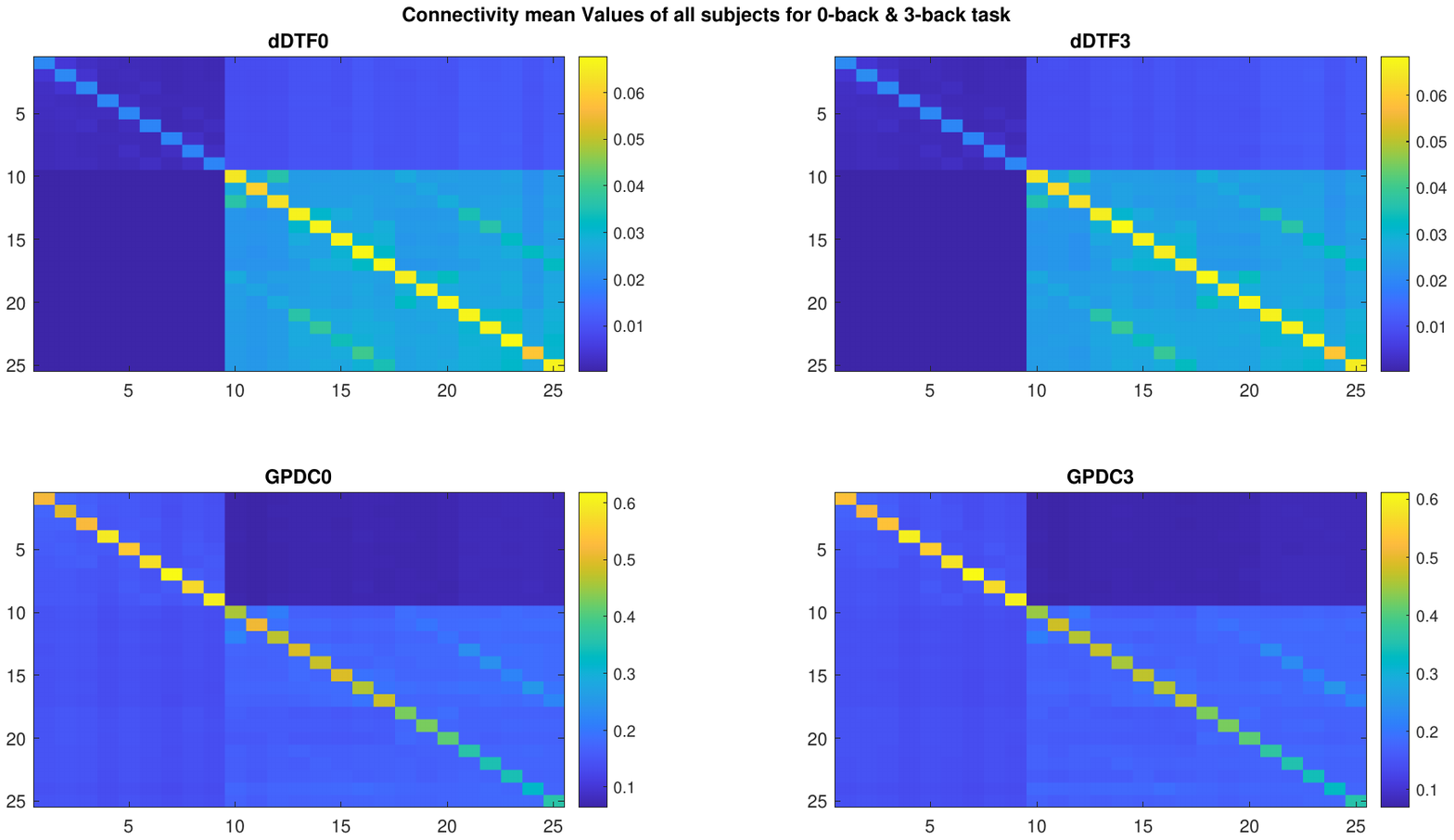} 
    \hspace{0.5cm} %
    \vspace{-0.4cm} \caption{Connectivity mean Values of all subjects for 0-back \& 3-back task. Top row: Mean dDTF (directed Directed Transfer Function) effective connectivity matrices across all 26 subjects for 0-back (dDTF0) and 3-back (dDTF3) tasks. Bottom row: Mean gPDC (generalized Partial Directed Coherence) effective connectivity matrices across all 26 subjects for 0-back (GPDC0) and 3-back (GPDC3) tasks. Color scales indicate the strength of effective connectivity, highlighting the directional information flow between regions (R1-R25).}
    \label{fig10}
\end{figure*}

The results of the dDTF and gPDC metrics for the 0-back and 3-back tasks, along with the average results across all participants, are presented in Figure~\ref{fig10}. Dark blue regions indicate minimal connectivity values close to zero, whereas lighter colors represent increased connectivity within the matrix. The dDTF, which measures the inflow of information, shows higher inflow values in regions R10, R11, ..., R25 compared to other areas. Additionally, the inflow from fNIRS data to EEG data was greater than that from EEG data to fNIRS data. The gPDC, which assesses the outflow of information, indicates that the outflow values in R10, R11, ..., R25 are lower than those in other regions. Moreover, the outflow from the EEG data to the fNIRS data is greater than that from the fNIRS data to the EEG data.

By comparing the dDTF and gPDC values, it was observed that the gPDC values were generally higher than the dDTF values, indicating that the outflow of information from EEG to fNIRS is greater than the inflow of information from fNIRS to EEG.

\begin{figure*}[h!]
    \centering
    \subfloat[0-back.]{\includegraphics[width=0.45\textwidth]{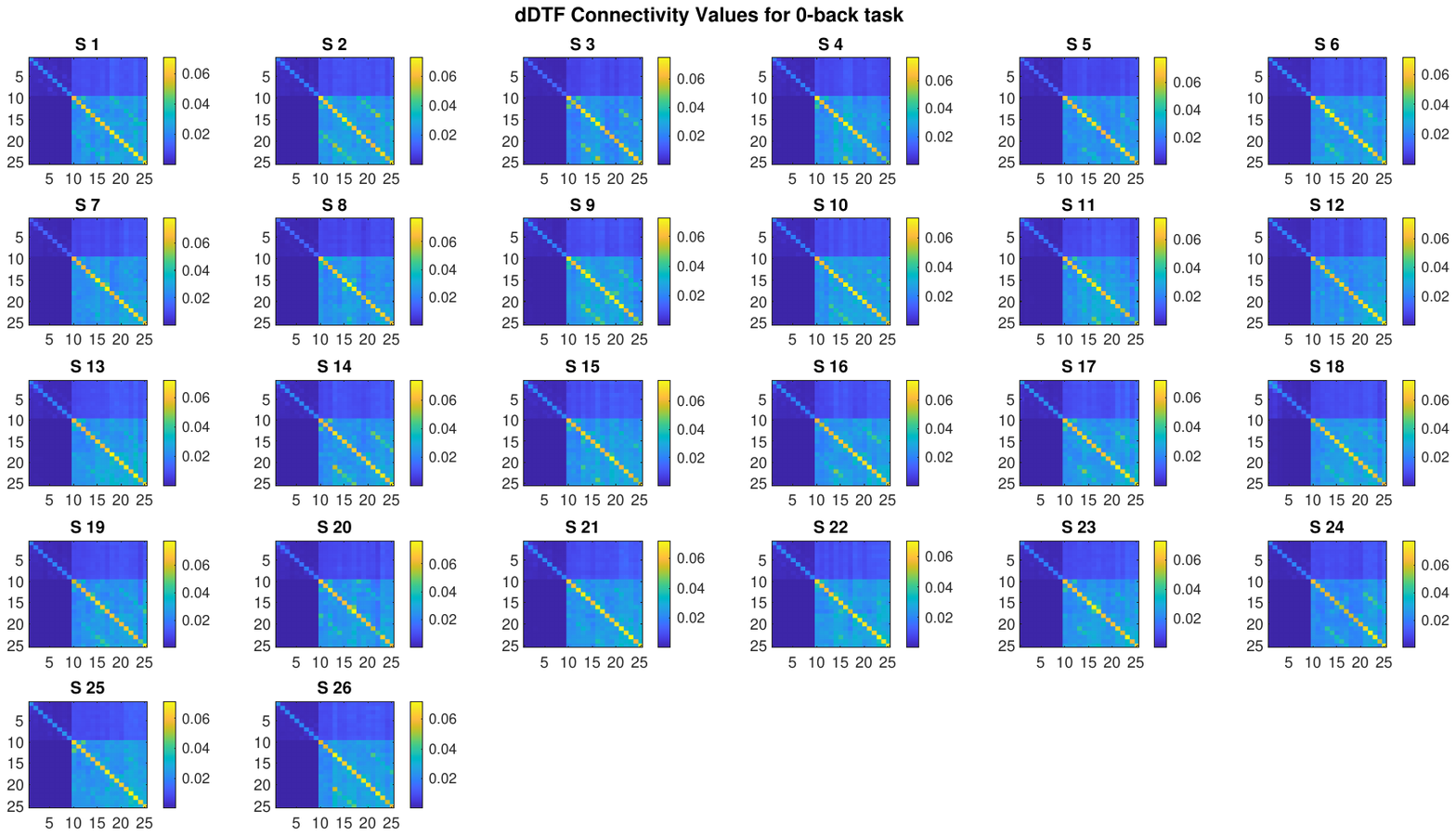}}
    \hfill
    \subfloat[3-back.]{\includegraphics[width=0.45\textwidth]{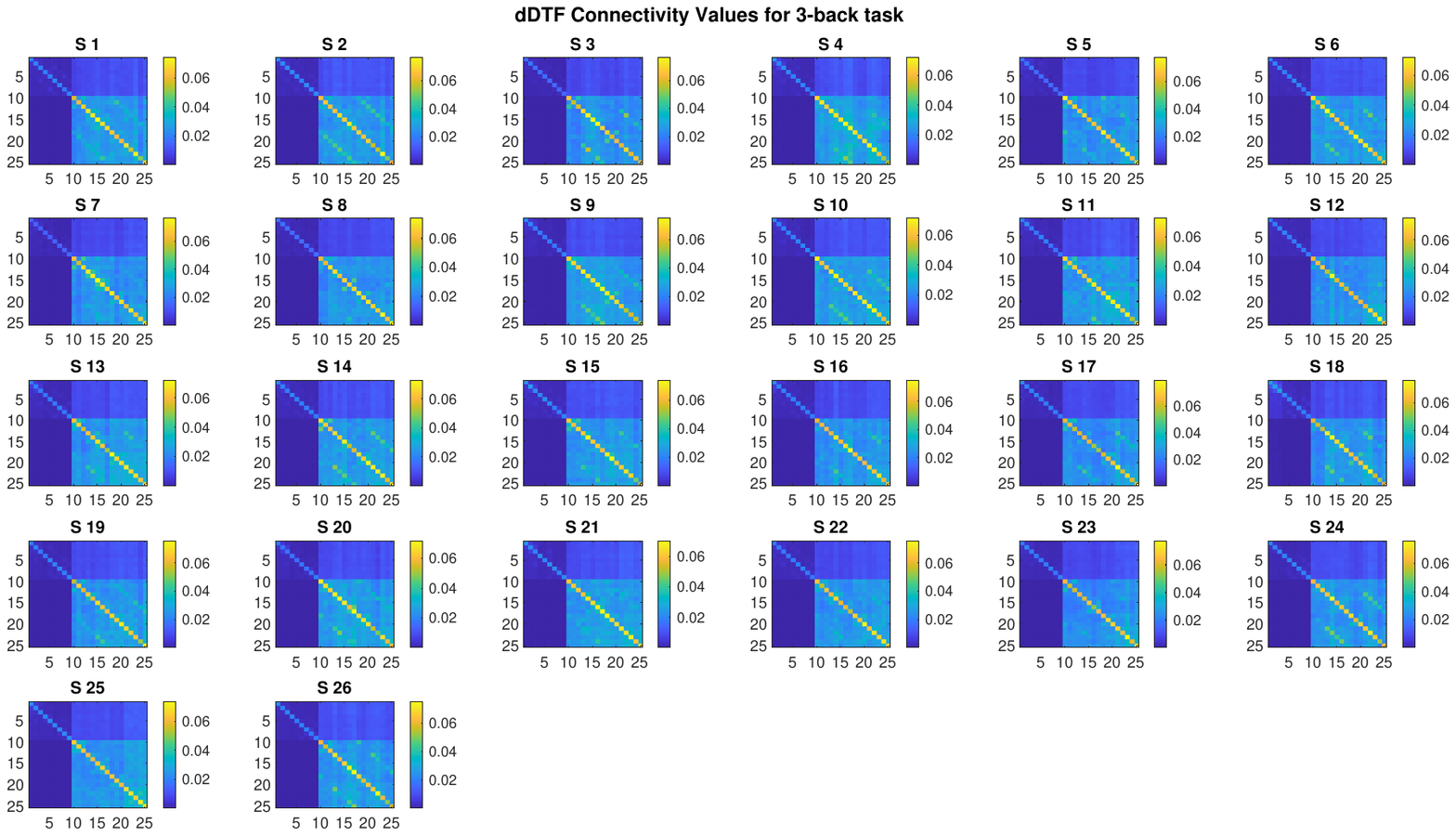}}
    \caption{directed Directed Transfer Function (dDTF) connectivity matrices for the 0-back (a) and 3-back (b) tasks. Each matrix represents the directional information flow between 25 brain regions (R1-R25) for 26 individual subjects (S1-S26), with data derived from combined EEG and fNIRS signals. The color scale, ranging from 0.02 to 0.06, indicates the strength of the directed connectivity. Higher values signify a stronger information inflow.}
    \label{fig11_12}
\end{figure*}

The dDTF metric quantifies the inflow of information from the various channels. For example, the inflow from fNIRS to EEG represents the proportion of information received by EEG from fNIRS relative to the total amount of information received by EEG.
As illustrated in Figure ~\ref{fig11_12}, a comparative analysis of the information exchange between EEG and fNIRS indicates that, across all participants, the inflow of information from fNIRS to EEG is consistently greater than the inflow from EEG to fNIRS.

\begin{figure*}[h!]
    \centering
    \subfloat[0-back.]{\includegraphics[width=0.45\textwidth]{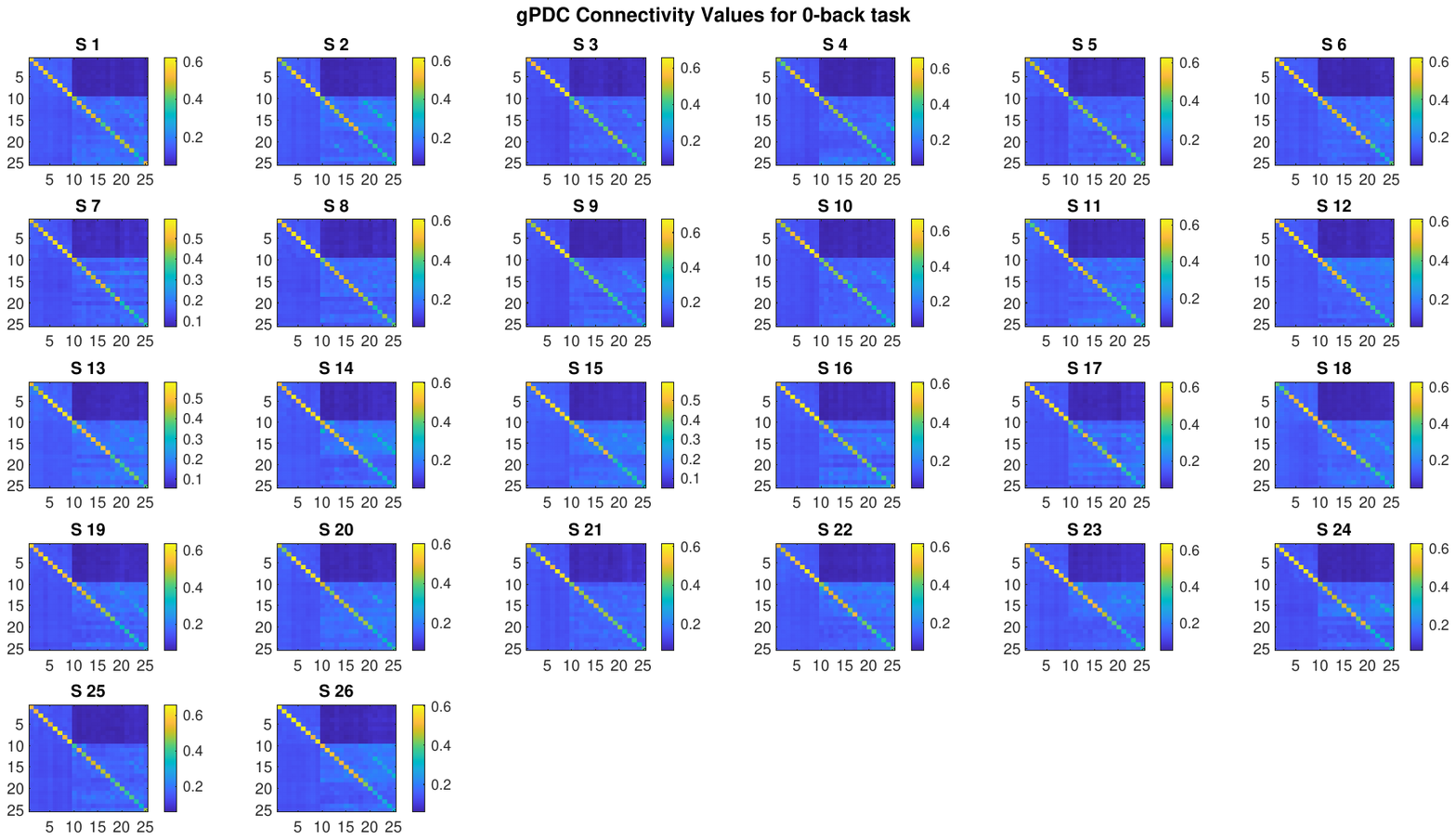}}
    \hfill
    \subfloat[3-back.]{\includegraphics[width=0.45\textwidth]{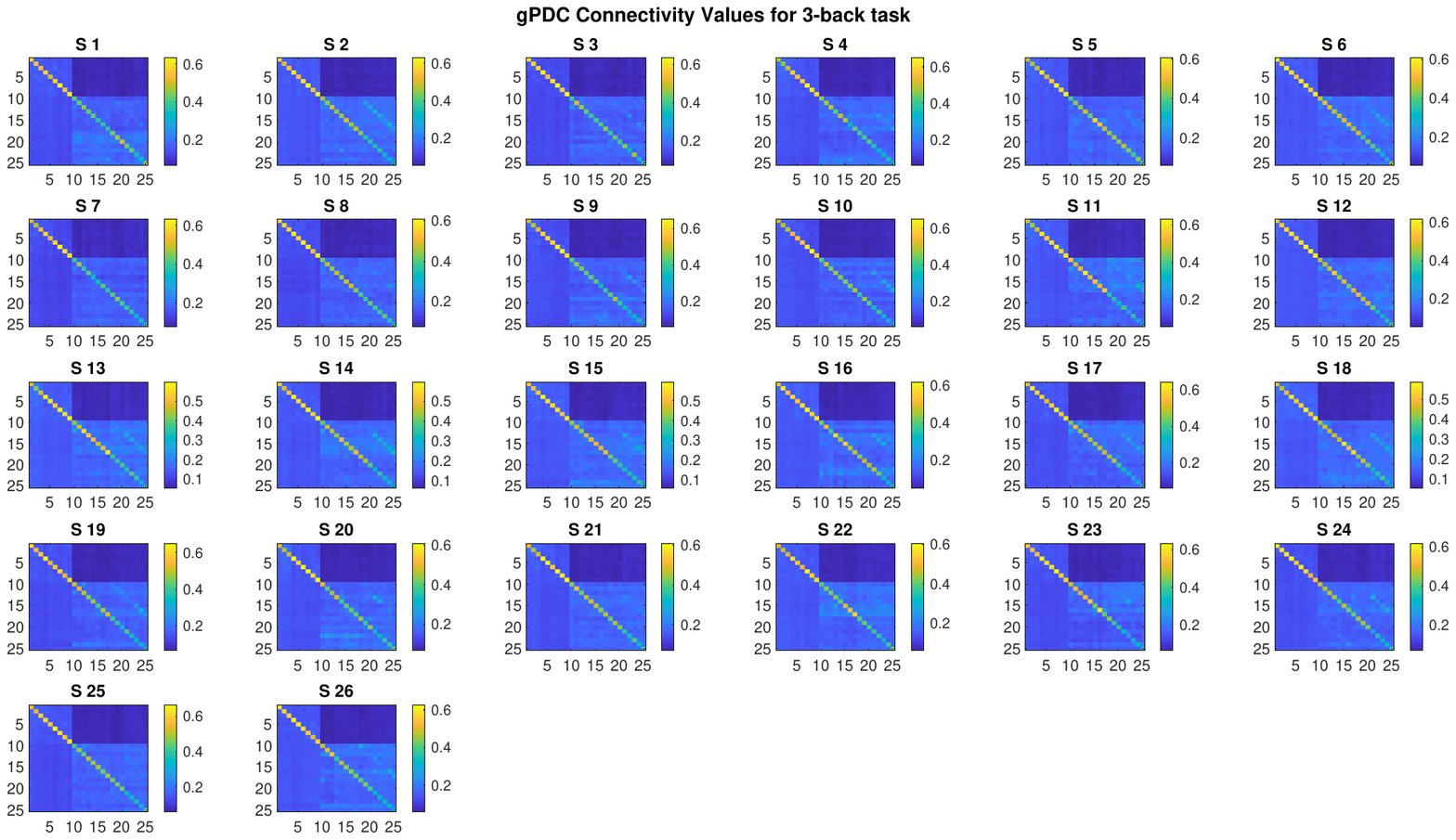}}
    \caption{generalized Partial Directed Coherence (gPDC) connectivity matrices for the 0-back (a) and 3-back (b) tasks. Each matrix represents the directional information flow between 25 brain regions (R1-R25) for 26 individual subjects (S1-S26), with data derived from combined EEG and fNIRS signals. The color scale, ranging from 0.1 to 0.6, indicates the strength of the directed connectivity. Higher values signify a stronger information outflow.}
    \label{fig13_14}
\end{figure*}

The gPDC metric quantifies the outflow of information  from the different channels. For instance, the  outflow from EEG to fNIRS represents the proportion of information transmitted from EEG to fNIRS relative to the total amount of information leaving the EEG signals. As illustrated in Figure~\ref{fig13_14}, a comparative analysis of the information flow between EEG and fNIRS indicates that, across all participants, the outflow of information from EEG to fNIRS is greater than the outflow from fNIRS to EEG. 

\subsection{Results of Statistical tests}
To assess the statistical significance of the differences in brain connectivity between the high and low mental workload groups, a Wilcoxon signed-rank test was conducted using SPSS. In this analysis, p-values less than 0.05 were considered indicative of a statistically significant difference.
The presented matrix displays the results of the Wilcoxon signed-rank test for comparing  brain connectivity between the two cognitive load conditions: low cognitive load (0-back) and high cognitive load (3-back). The rows and columns of this matrix correspond to different EEG and fNIRS brain regions, categorized as R1–R25.  

The p-values in each cell of the matrix indicate the statistical significance of the differences in connectivity between these regions under the two cognitive load conditions. Darker colors (blue) represent p-values less than 0.05, signifying a statistically significant difference in brain connectivity between the 0-back and 3-back conditions. In contrast, the white cells indicate non-significant differences between these conditions.  

Overall, these findings highlight the changes in brain connectivity patterns in response to increased cognitive load, particularly in specific brain regions that may play a more critical role in cognitive task processing.
The results of statistical tests for each brain connectivity measure are discussed in detail in the following sections.

\subsubsection{Statistical tests for MSC:}
\begin{figure*}[h!]
    \centering
 \hspace{-1.5cm}
  \includegraphics[scale=0.55]{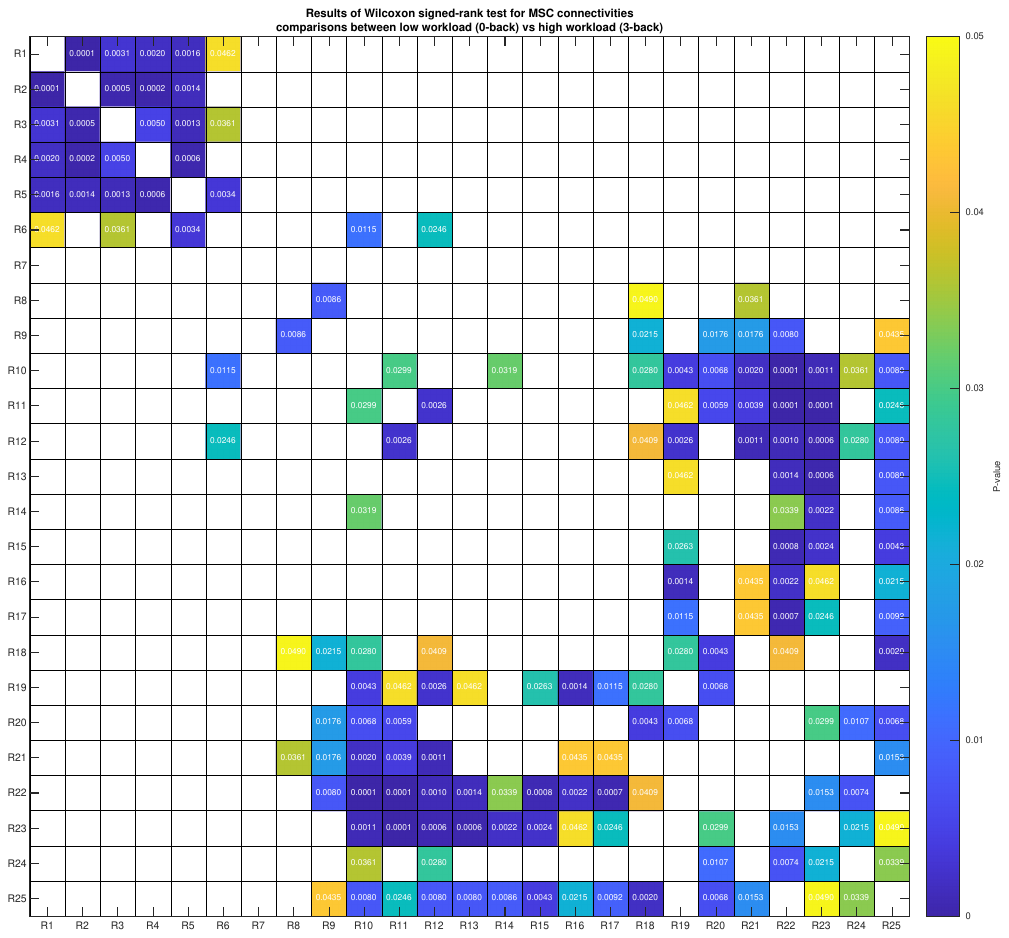}
    \vspace{-1cm}
    \caption{Results of Wilcoxon signed-rank test for Magnitude Squared Coherence (MSC) connectivities comparing low workload (0-back) versus high workload (3-back) conditions. The matrix displays p-values for statistical significance of differences in MSC connectivity between 0-back and 3-back tasks across 25 regions (R1-R25). The color scale represents p-values, with darker blue indicating statistically significant difference and lighter colors/white indicating less significant differences. Regions R1-R9 correspond to EEG regions, R10-R17 to OXY regions, and R18-R25 to DEOXY regions.}
    \label{fig15}
\end{figure*}

Based on Figure~\ref{fig15}, which shows the results of the Wilcoxon signed-rank test for MSC connectivity under two levels of cognitive load, significant changes were observed across different brain regions. it can be concluded that an increase in cognitive load from the 0-back to the 3-back task extensively impacts brain coherence. This effect is particularly statistically significant in the EEG-related frontal and central brain regions, indicating an increase in coherence within networks responsible for executive functions and working memory. Furthermore, statistically significant differences are also observed in both DEOXY and OXY, which confirms the impact of cognitive load on fNIRS as well.

\subsubsection{Statistical tests for PLV:}
\begin{figure*}[h!]
    \centering
\hspace{-1.5cm}
  \includegraphics[scale=0.55]{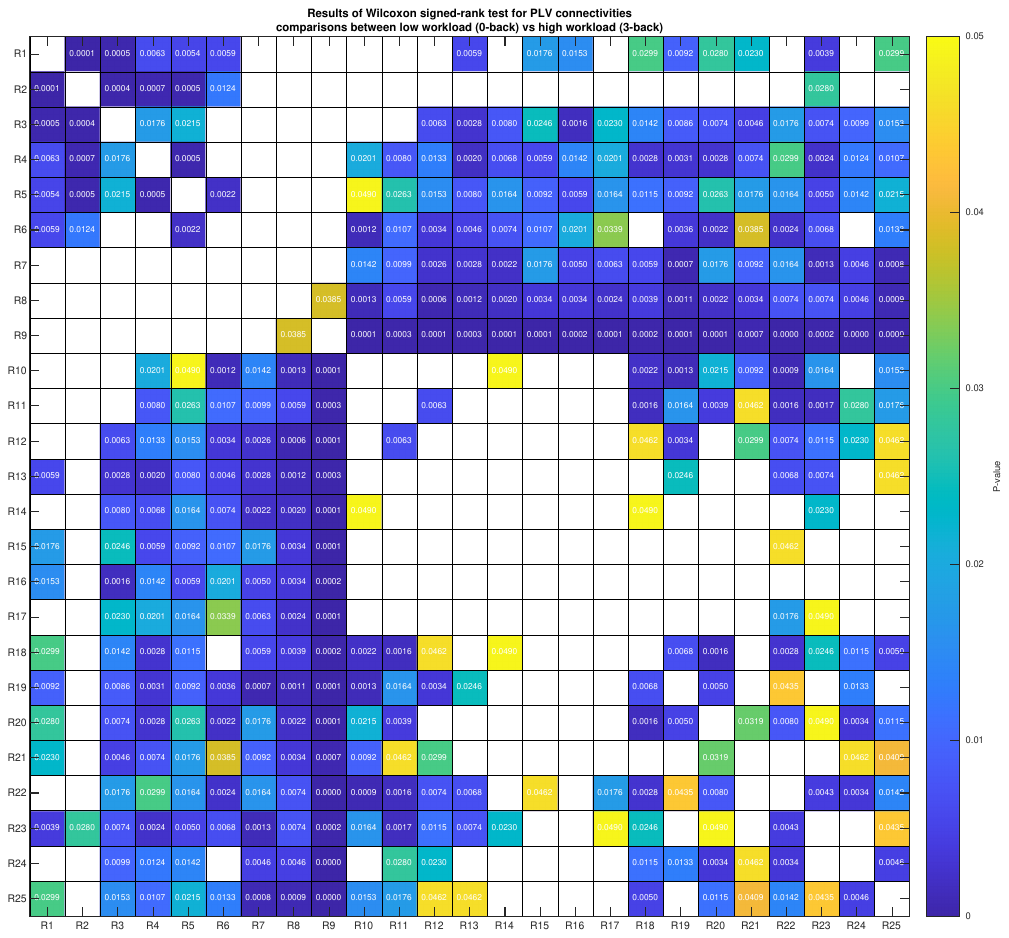}
    \vspace{-1cm}
    \caption{Results of Wilcoxon signed-rank test for Phase Locking Value (PLV) connectivities comparing low workload (0-back) versus high workload (3-back) conditions. The matrix displays p-values for statistical significance of differences in PLV connectivity between 0-back and 3-back tasks across 25 regions (R1-R25). The color scale represents p-values, with darker blue indicating statistically significant difference and lighter colors/white indicating non-significant differences. Regions R1-R9 correspond to EEG regions, R10-R17 to OXY regions, and R18-R25 to DEOXY regions.}
    \label{fig16}
\end{figure*}

Based on Figure~\ref{fig16}, the analysis of PLV changes in the EEG, OXY, and DEOXY regions revealed distinct patterns in response to an increased cognitive load. it can be concluded that an increase in cognitive load from the 0-back to the 3-back task extensively impacts the temporal synchrony of brain activity. This effect is particularly statistically significant in the EEG-related frontal and central brain regions, indicating an increase in synchrony within areas responsible for executive functions and working memory. Statistically significant differences are also observed in the temporal coordination of fNIRS (R10 to R25), which confirms the impact of cognitive load on both OXY and DEOXY. Furthermore, an increase in synchrony in both EEG and fNIRS networks is observable in response to the higher cognitive demand.

\subsubsection{Statistical tests for PCC:}
\begin{figure*}[h!]
    \centering
\hspace{-1.5cm}
  \includegraphics[scale=0.55]{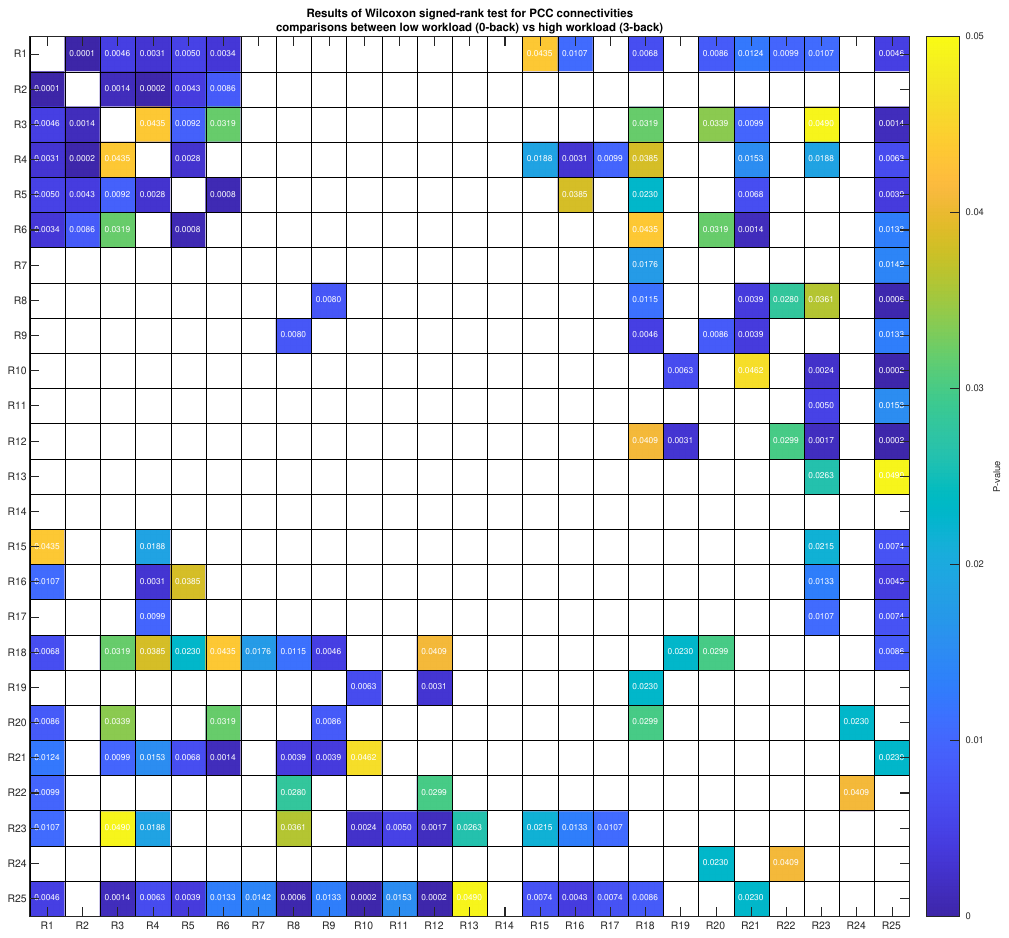}
    \vspace{-1cm}
    \caption{Results of Wilcoxon signed-rank test for Pearson Correlation Coefficient (PCC) connectivities comparing low workload (0-back) versus high workload (3-back) conditions. The matrix displays p-values for statistical significance of differences in PCC connectivity between 0-back and 3-back tasks across 25 regions (R1-R25). The color scale represents p-values, with darker blue indicating statistically significant difference and lighter colors/white indicating non-significant differences. Regions R1-R9 correspond to EEG regions, R10-R17 to OXY regions, and R18-R25 to DEOXY regions.}
    \label{fig17}
\end{figure*}

Based on Figure~\ref{fig17}, the statistical analysis of PCC-based brain connectivity, it can be concluded that an increase in cognitive load from the 0-back to the 3-back task significantly impacts brain connectivity. This effect is particularly statistically significant in the EEG-related frontal and central brain regions, as well as in the DEOXY-related frontal regions. These areas, which are responsible for executive functions and working memory, increase their coordination and connectivity to respond to the higher cognitive demand. These changes are also observed in the connectivity between the EEG and fNIRS regions. In contrast, a statistically significant difference is not seen in the connectivity within the OXY regions. This finding indicates that the increase in cognitive load has not significantly altered the internal connectivity patterns of this network.

\subsubsection{Statistical tests for dDTF and gPDC:}
The results of the Wilcoxon signed-rank test for effective connectivity (dDTF and gPDC) indicate a highly significant difference (\( p = 8.3 \times 10^{-6} \)) in information flow between EEG and fNIRS, confirming that connectivity values in the EEG-to-fNIRS and fNIRS-to-EEG directions are statistically distinct. This suggests an asymmetric neurovascular interaction, where EEG activity predominantly drives fNIRS responses, aligning with neurovascular coupling models where electrical activity induces metabolic demand, leading to vascular adaptation. The weaker fNIRS-to-EEG connectivity implies that hemodynamic changes do not exert an equivalent immediate effect on neural oscillations, reinforcing the feedforward dominance of EEG over fNIRS in cognitive workload processing. These findings highlight the necessity of directional analysis in multimodal neuroimaging, demonstrating that effective connectivity metrics (dDTF, gPDC) provide critical insights into neural-vascular interactions, with implications for mental workload assessment and cognitive neuroscience research.

\section{Discussion}

 This study provides a comprehensive investigation into brain connectivity and information flow during mental workload by integrating EEG and fNIRS signals, utilizing the N-back task to induce varying cognitive demands. Our findings provide novel insights into the reorganization of brain networks under workload and crucially, elucidate the directional interplay between neural and hemodynamic responses.

Our analysis of functional connectivity  revealed consistent patterns across PCC, PLV and MSC metrics. Specifically, we observed a significant increase in brain connectivity in frontal regions (R1, R2, R3) as mental workload increased from the 0-back to the 3-back condition. This heightened connectivity in frontal areas, well-known for their roles in working memory and cognitive control, aligns with the expected neural recruitment necessary to manage increased task complexity.

The most critical contribution of this study lies in the effective connectivity  analysis, employing dDTF and gPDC. Our results demonstrated a highly significant directional information flow from EEG to fNIRS (\( p = 8.3 \times 10^{-6} \)). This finding powerfully substantiates the concept of neurovascular coupling, providing empirical evidence that neural electrical activity, as captured by EEG, predominantly drives the subsequent hemodynamic responses measured by fNIRS during cognitive processing under workload. The consistent observation that outflow from EEG to fNIRS was greater than inflow from fNIRS to EEG, while inflow from fNIRS to EEG was generally greater than EEG to fNIRS, indicates a strong feedforward dominance of neural activity influencing vascular changes. This supports the widely accepted notion that neuronal activation triggers metabolic demands, leading to local changes in cerebral blood flow and oxygenation, which fNIRS then detects. This directional insight is paramount for understanding the fundamental mechanisms by which the brain adapts to cognitive effort, moving beyond mere statistical associations to infer causal influence. Our findings resonate with previous multimodal studies that have explored brain-heart interactions and EEG-PPG coupling, further emphasizing the importance of directional analysis in unraveling complex physiological interdependencies.

Compared to prior single-modality studies, our integrated approach provides a more nuanced understanding of MWL-related connectivity. For instance, previous EEG-based studies have reported decreased characteristic path lengths and increased global efficiency in high workload conditions, suggesting functional integration of brain networks. Similarly, fNIRS-only studies have shown increased connectivity in prefrontal cortex regions under elevated cognitive load. However, these unimodal approaches do not capture the directionality of neurovascular interactions. By integrating EEG and fNIRS and applying effective connectivity analysis, our findings fill this gap, offering a new dimension to MWL assessment.

Our study primarily focused on investigating changes in brain connectivity patterns and analyzing information flow and did not involve classification. In future work, the features and insights derived from this study can be leveraged for classification tasks, potentially incorporating advanced methods such as deep learning. Further research could also explore connectivity analysis in additional frequency bands, delve deeper into graph theoretical approaches, or apply these findings to real-time classification scenarios.

\section{Conclusions}
This study successfully integrated EEG and fNIRS to investigate brain connectivity and information flow under varying levels of mental workload induced by the N-back task. Our analysis of functional connectivity, using PCC, PLV, and MSC, revealed increased connectivity, particularly in frontal regions, as mental workload increased from the 0-back to the 3-back condition. Crucially, the effective connectivity analysis employing dDTF and gPDC demonstrated a statistically significant directional information flow from EEG to fNIRS. This finding strongly suggests that neural electrical activity, as measured by EEG, exerts a dominant influence on the hemodynamic responses captured by fNIRS during cognitive processing under workload. The highly significant p-value obtained for both dDTF and gPDC underscores this directional influence and supports the concept of neurovascular coupling, in which neuronal activity drives subsequent vascular changes. These results highlight the synergistic benefits of EEG-fNIRS integration for a comprehensive understanding of brain network dynamics and information flow in response to mental workload. The observed directional connectivity provides valuable insights into the neurophysiological mechanisms underlying cognitive effort and has implications for developing more robust and nuanced methods for real-time mental workload monitoring and assessment in various applied settings.

\vspace{0.5cm}

\textbf{CRediT authorship contribution statement}
\vspace{0.25cm}
\noindent 

\textbf{Mohaddese Qaremohammadlou:} Conceptualization, Formal analysis, Investigation, Methodology, Software, Validation, Visualization, Writing – original draft, Writing – review \& editing. \textbf{Mohammad Bagher  Shamsollahi:} Conceptualization, Investigation, Methodology, Supervision, Validation, Writing – review \& editing.
\vspace{0.5cm}

\textbf{ Declaration of Competing interest}

\vspace{0.25cm} 

It is declared that no known competing financial interests or personal relationships could have appeared to influence the work reported in this paper.
\vspace{0.5cm}

\textbf{ Funding sources}

\vspace{0.25cm} 

This research did not receive any specific grant from funding agencies in the public, commercial, or not-for-profit sectors.
 \vspace{0.5cm}

\textbf{ Data availability}

\vspace{0.25cm} 

The data is available for download from the following link: \href{http://doc.ml.tu-berlin.de/simultaneous_EEG_NIRS/. }{ doc.ml.tu-berlin.de/simultaneous EEG NIRS/.} Further details and comprehensive information regarding this dataset are provided in the corresponding publication \cite{shin2018simultaneous}.


\bibliographystyle{IEEEtran}
\bibliography{jour}



\end{document}